\documentclass[11pt]{article}
\usepackage{amsfonts}
\usepackage{amsmath,amssymb}
\usepackage{bbm}
\usepackage{bm}
\usepackage{color}
\usepackage{slashed}
\usepackage{cite}
\usepackage[utf8]{inputenc}
\usepackage[toc,page]{appendix}
\usepackage{graphicx}
\usepackage{placeins}
\usepackage{caption}
\usepackage{subfigure}
\usepackage{booktabs}
\usepackage{comment}
\usepackage{physics}
\usepackage{float}
\usepackage{bbold}
\DeclareGraphicsExtensions{.pdf,.png,.jpg}

\usepackage[colorlinks=false,citecolor=cyan,urlcolor=blue,bookmarks=true,bookmarks=true,bookmarksopen=true,bookmarksnumbered=true,bookmarksopenlevel=3]{hyperref}

\definecolor{airforceblue}{rgb}{0.36, 0.54, 0.66}
\definecolor{steelblue}{rgb}{0.27, 0.51, 0.71}
\definecolor{amber}{rgb}{1.0, 0.49, 0.0}
\definecolor{darkgreen}{rgb}{0.0, 0.5, 0.0}
\definecolor{amber}{rgb}{1.0, 0.49, 0.0}

\DeclareMathAlphabet{\mathpzc}{OT1}{pzc}{m}{it}

\allowdisplaybreaks[1]
\def\simg{{\ \lower-1.2pt\vbox{\hbox{\rlap{$>$}\lower6pt\vbox{\hbox{$\sim$}}}}\ }}
\def\siml{{\ \lower-1.2pt\vbox{\hbox{\rlap{$<$}\lower6pt\vbox{\hbox{$\sim$}}}}\ }}

\makeatletter \@addtoreset{equation}{section} \makeatother

\newcommand*{\diff}[1]{\text{d}#1}

\newcommand*{\der}[2]{\frac{d #1}{d #2}}

\begin{document}

\flushbottom

\begin{titlepage}

\begin{centering}

\vfill

{\Large{\bf
Dark matter interacting via a massive spin-2 mediator \\in warped extra-dimensions}}


\vspace{0.8cm}

A.~de~Giorgi$^{\rm a\,}$\footnote{arturo@mpp.mpg.de}
and S.~Vogl$^{\rm b,a\,}$\footnote{ stefan.vogl@physik.uni-freiburg.de}

\vspace{0.8cm}

{\em $^{\rm{a}}$Max-Planck-Institut 
f\"ur Physik (MPP)\\ F\"ohringer Ring 6,
80805 M\"unchen,
Germany} 
\\
\vspace{0.15 cm}
{\em $^{\rm{b}}$Albert-Ludwigs-Universität Freiburg, Physikalisches Institut\\
Hermann-Herder-Str. 3, 79104 Freiburg,
Germany}
\vspace*{0.8cm}

\end{centering}
\vspace*{0.3cm}
 
\noindent

\textbf{Abstract}:  We study dark matter interacting via a massive spin-2 mediator. To have a consistent effective theory for the spin-2 particle, we work in a warped extra-dimensional model such that the mediator(s) are the Kaluza-Klein (KK) modes of the 5D graviton. We pay close attention to dark matter annihilations into KK-gravitons. Due to the high energy behavior of longitudinal modes of spin-2 fields, these channels exhibit a tremendous growth at large center of mass energies $\sqrt{s}$ if only one spin-2 mediator is considered. For the first time, we include the full KK-tower in this dark matter production process and find that this growth is unphysical and cancels once the full field content of the extra-dimensional theory is taken into account. Interestingly, this implies that it is not possible to approximate the results obtained in the full theory with a reduced set of effective interactions once $\sqrt{s}$ is greater than the first graviton mass. This casts some doubt on the universal applicability of previous studies with spin-2 mediators within an EFT framework and prompts us to revisit the phenomenological allowed parameter space of gravitationally interacting scalar dark matter in warped extra-dimensions.

\vfill

\end{titlepage}
\setcounter{footnote}{0}

\begin{section}{Introduction}

Over the last decades, a stupendous amount of evidence for the presence of Dark Matter (DM) in our Universe has been collected. This observational program culminates in the precision measurement of the CMB anisotropies by the Planck collaboration  that allows a percent level determination of its abundance \cite{Akrami:2018vks}. However, all these observations only test the gravitational effect of DM and despite large efforts, a detection of other interactions of the DM with the Standard Model (SM) is still elusive. As interactions mediated by SM particles and popular SM extensions are increasingly constrained, see e.g. \cite{Arcadi:2019lka,Escudero:2016gzx,Blanco:2019hah,Duerr:2016tmh},
alternative ideas are hotly sought after. One possibility that has attracted more attention in recent years is that DM interacts with the SM via a massive spin-2 mediator \cite{Lee:2013bua,Lee:2014caa,Rueter:2017nbk,Folgado:2019gie,Folgado:2019sgz,Babichev:2016bxi,Albornoz:2017yup,Bernal:2018qlk,Kraml:2017atm,Bernal:2020fvw}\footnote{Interactions mediated by the graviton, i.e. the massless spin-2 field of quantized general relativity, are also a possibility \cite{Garny:2015sjg,Ema:2018ucl}.}. To date, there is no renormalizable model for the origin and nature of a massive spin-2 particle. Therefore, most studies either directly follow a simplified effective field theory approach or consider an origin from gravity in compactified extra-dimension, see however \cite{Babichev:2016bxi,Albornoz:2017yup}.

In these models, an important complication arises once external spin-2 particles are considered. Similar to the longitudinal modes of massive vector bosons the longitudinal part of the polarization vectors of massive spin-2 fields, $\epsilon^{\mu \nu}_0$,  exhibit an enhancement for large energies with $\epsilon^{\mu \nu}_0\propto E^2/m^2$ where $E$ is the energy and $m$ the mass of the spin-2 particle $G$. In a theory with a single massive tensor field, the amplitude for the pair production of $G$ from matter fields grows as $E^6$ in the high energy limit. This leads to a strong growth of the cross section $\propto E^{10}$ which has a major impact on DM production in the early Universe if processes with $\sqrt{s}\geq m_G$ are relevant. Depending on the production mechanism the parameter range in which this is the case changes. On the one hand, for DM produced via thermal freeze-out, this high energy enhancement is important if its mass $m_{\tiny{DM}}$ exceeds the mass of the mediator, see e.g. \cite{Lee:2014caa,Folgado:2019gie}. On the other hand, for a freeze-in scenario, the relevant quantity is the temperature $T$ instead and a strong increase of the production rate for $T\geq m_G$ was found in e.g.~\cite{Bernal:2018qlk,Bernal:2020fvw}. 

Such a large growth of the amplitudes heralds a breakdown of the effective theory at energies well below the explicit scale $\Lambda$ that enters in the effective operator. This is not necessarily problematic in a general effective theory and merely indicates that the unspecified UV completion has to become relevant at lower scales than naively expected. However, it is surprising in extra-dimensional models since they ought to remain perturbative up to the scale of higher dimensional gravity. These clashing expectations can be resolved by realizing that the 4D theory that arises from the KK-decomposition of extra dimensions does not just contain a single spin-2 field but a tower of KK-gravitons and their interactions instead. We recently analyzed the high energy limit of KK-graviton production in warped extra-dimensions and found that the contribution of the other KK-gravitons conspire to cancel the contributions that grow faster than $E^2$ \cite{deGiorgi:2020qlg} \footnote{A related issue has also received attention in the context of spin-2 scattering. In a theory with a single massive graviton, the amplitude of $G G\rightarrow G G$ grows like $s^5$. It has been shown that the tree-level exchange of any number of spin-0 and spin-1 fields cannot remove this behavior\cite{Bonifacio:2019mgk} but in extra-dimensional theories the growth is reduced to $E^2$ once the full KK-tower of massive gravitons is included, see \cite{Bonifacio:2019ioc} or \cite{Chivukula:2020hvi,Chivukula:2019rij,Chivukula:2019zkt}. }.
 Naturally, this will have an impact on the DM production rates and can change the amount produced in the early Universe considerably. Therefore, it is of great interest to revisit the production of DM via spin-2 mediators in the context of extra-dimensions.
For concreteness, we work with the warped extra-dimensions model of Randall and Sundrum 
\cite{Randall:1999ee}. 
We expect that our qualitative conclusions regarding the high energy behavior of the DM annihilation (or production) rates will also hold in extra-dimensional models with a different geometry. The results of our phenomenological study are clearly limited to the model we consider.

The structure of this paper is as follows. First, we introduce the Randall-Sundrum (RS) model which acts as our blueprint for spin-2 mediators from higher dimensional gravity. In Sec.~\ref{sec:DM_production} we discuss the production of DM in the early Universe focusing on the freeze-out mechanism. We pay particular attention to the modifications of the annihilation cross sections that arise once the full KK-tower is considered. Next, we reassess the phenomenology of thermally produced DM in the Randall-Sundrum model. Finally, we present our conclusions in Sec.~\ref{sec:Conclusions}. Additional material concerning the decay width of the graviton and radion and the derivation of some relations between the couplings of the KK-gravitons is presented in two Appendices. 

\end{section}

\begin{section}{The Randall-Sundrum model}
\label{sec:RS_model}

The RS model consists of a 5D space compactified under a $S^1/\mathbb{Z}^2$ orbifold symmetry \cite{Randall:1999ee}. After compactification, one finds a  5D bulk space bounded by two branes. Only gravity propagates in the 5th dimension while SM fields are localized on one of the branes.
To allow for a DM candidate, we extend the model minimally and add a single, only gravitationally interacting scalar that is localized on the same brane as the SM.   In the following, we will briefly summarize the relevant ingredients of the RS model. A more pedagogical and detailed introduction can for example be found in \cite{Rattazzi:2003ea,Kribs:2006mq,Raychaudhuri:2016kth}.

The total action of the model is
\begin{equation}
        S = S_{\text{bulk}}+S_{\text{UV}}+S_{\text{IR}} \, ,
\end{equation}
and includes a contribution from the bulk and the UV- and IR-branes, respectively. 
Only gravity propagates in the fifth dimension and, therefore, the bulk contribution reads
    \begin{align}
    \label{eq:RS_Action}
            & S_{\text{bulk}} = \frac{1}{2}M_5^3\int d^4x \int\limits_{-\pi}^\pi d\varphi \sqrt{G}(R-2\Lambda_B ) 
    \end{align}
where $M_5$ is the Planck mass in 5D while $G$ is the determinant of the higher dimensional metric $G_{MN}$ and $R$ denotes the Ricci scalar. 
We use large Latin letters, e.g. $M,N$, to indicate indices that run over all five dimensions while small Greek letters, e.g. $\mu,\nu$, only run over the usual four dimensions. 
The coordinate of the full 5D space, therefore, reads $x^M=(x^\mu,y)$.  In the equation above, $y$ has been replaced by a dimensionless coordinate $\varphi=y/r_c$ where $r_c$ is a measure of the size of the extra-dimension. 
To get the desired phenomenology, a bulk vacuum energy term $\Lambda_B$ has to be included. The contributions of the two branes  located at $y=0$ and $y=\pi r_c$, respectively, are given by 
            \begin{align}
            & S_{\text{UV}} =\int d^4x\int\limits_{-\pi}^\pi d\varphi \sqrt{-g_{\text{UV}}}(-V_{\text{UV}}+\mathcal{L}_{\text{UV}})\delta(\varphi) \nonumber \ , \\
            & S_{\text{IR}}=\int d^4x\int\limits_{-\pi}^\pi d\varphi \sqrt{-g_{\text{IR}}}(-V_{\text{IR}}+\mathcal{L}_{\text{IR}})\delta(\varphi-\pi) \ ,
\end{align}
where $g_{\text{UV/IR}}$ are the metrics induced by $G$ on the branes, $\mathcal{L}_{\text{UV/IR}}$ indicate the corresponding matter fields Lagrangians and $V_{\text{UV/IR}}$ parametrize the branes' vacuum energies. We consider the case where both the SM and the DM are localized on the IR brane, i.e. $\mathcal{L}_{\text{IR}}=\mathcal{L}_{\text{SM}}+ \mathcal{L}_{\text{DM}}$, while $\mathcal{L}_{\text{UV}}=0$ for simplicity. To keep our  model minimal we introduce a single scalar $\phi$ that only interacts gravitationally such that
\begin{align}
\mathcal{L}_{\text{DM}}=\frac{1}{2} \partial_\mu \phi \partial^\mu \phi -\frac{1}{2} m^2 \phi^2\ .
\end{align}  

Solving Einstein's equation in the absence of matter and choosing the vacuum energy contributions such that 4D Poincaré invariance is realized on the branes leads to an infinitesimal line element
\begin{equation}
    d s^2= e^{-2k |y|}\eta_{\mu\nu} d x^\mu dx^\nu- d y^2 \ ,
\end{equation}
where $\eta_{\mu \nu}$ is the Minkowski metric in 4D and the warping parameter $k \equiv \sqrt{\frac{-\Lambda_B}{6}}$. It is often convenient to work with dimensionless quantities instead and one defines $\mu= k r_c$.

We want to work in an effective field theory in 4D which captures the low energy behavior of the model outlined above. Three steps are required to derive this EFT. First, one performs
a weak-field expansion of the 5D metric in terms of an expansion parameter $\kappa =2/M_5^{3/2}$, that leads to a Lagrangian involving a tensor perturbation $\hat{h}_{\mu \nu}$, i.e. the 5D graviton, and a scalar perturbation $\hat{r}$, the so-called radion. A detailed discussion of this expansion and the resulting Lagrangian up to third order in the fields can for example be found in \cite{deGiorgi:2020qlg,Chivukula:2020hvi}.  Second, one utilizes a Kaluza-Klein decomposition of the  5D graviton
\begin{align}
        \hat{h}_{\mu\nu}(x,y)&=\sum\limits_{n=0}^\infty\frac{1}{\sqrt{r_c}}h^{(n)}_{\mu\nu}(x)\ \psi_n(\varphi(y))
\end{align}
to express it in terms of a tower of 4D spin-2 fields $h^{(n)}_{\mu \nu}(x)$, which do not depend on the 5th dimension, and the wave functions $\psi_n(y)$ along the 5th dimension. 
The wave functions are fixed by a differential equation~\cite{Davoudiasl:1999jd}
\begin{equation}
\label{eq:mass-phi}
    \frac{1}{r_c^2} \der{}{\varphi}\left[A(\varphi)^4\der{\psi_n}{\varphi}\right]=-m_n^2 A^2 \psi_n \ ,
\end{equation}
where $m_n$ is the mass of the $n$-th graviton and  $A(\varphi)=e^{-\mu|\varphi|}$.
A successful description of 4D gravity demands that the lightest graviton is massless and fixes the normalization of the zero mode $\psi_0 = \sqrt{\frac{\mu}{1-e^{-2\mu\pi}}}$.
The other masses and the associated wave functions are obtained by solving eq.~\ref{eq:mass-phi} with boundary conditions $\partial_\varphi\psi_n|_{\varphi=0,\pi}=0$. 
The general solutions can be expressed in terms of Bessel-$J$ and Bessel-$Y$ functions and read
\begin{equation}
    \psi_n(\varphi)=\frac{e^{2\mu|\varphi|}}{N_n}\left[J_2(z_n)+\alpha_n Y_2(z_n) \right] \quad \mbox{where} \quad z_n \equiv m_n \frac{e^{\mu|\varphi|}}{k} \ .
\end{equation}
The normalization constants $N_n$ are fixed by the scalar product
\begin{equation}
\label{eq:orthogonality}
      \left<\psi_n,\psi_m \right> =\int\limits_{-\pi}^{\pi} d\varphi \ A(\varphi)^2 \ \psi_n(\varphi)\psi_m(\varphi)=\delta_{n,m} \ .
\end{equation}
For $e^{-\mu \pi} \ll 1$ the second term can be neglected and the solutions simplify to
\begin{equation}
     \psi_n(\varphi)\simeq \frac{e^{2\mu |\varphi|}}{N_n}J_2\left(\gamma_n e^{\mu(|\varphi|-\pi)}\right) \quad \mbox{and}\quad   m_n \simeq k \gamma_n e^{-\mu\pi} \ ,
\end{equation}
where  $\gamma_n$ denotes the $n$th zero of $J_1(x)$.

For convenience it is conventional to perform a similar change in the normalization of the radion 
\begin{equation}
    \hat{r}(x)=\frac{1}{\sqrt{r_c}} \psi_r \ r(x) \
\end{equation}
even though the $\hat{r}$ field does not possess a $y$-dependence. Now we have a Lagrangian in which the fields $h_{\mu \nu}$, $r$ are already 4-dimensional but we still need to perform the integration over the 5th dimensions to arrive at a fully 4D theory. This last step fixes the coefficients of the interactions in terms of integrals over the 5D wave functions.

The leading term in the interaction between gravitons and matter reads
\begin{equation}
    \mathcal{L}_{\text{int}}^{(1)}=-\frac{1}{2}\kappa T_{\mu\nu}\hat{h}^{\mu\nu}(x,\varphi=\pi)=-\frac{1}{2}\kappa T_{\mu\nu}\left(\sum\limits_{n=0}^\infty \frac{1}{\sqrt{r_c}}h_n^{\mu\nu}(x)\psi_n(\pi)\right) \ ,
\end{equation}
where $T_{\mu\nu}$ is the energy-momentum tensor of the matter field.
Matching the massless graviton contribution to the equivalent term in GR fixes the relation between the (reduced) Planck mass in 4D, $M_{Pl}$ and the parameters of the 5D theory 
\begin{equation}
    \frac{1}{2}\kappa \frac{1}{\sqrt{r_c}}\psi_0 =\frac{1}{M_{Pl}}  .
\end{equation}
Plugging in explicit expressions this leads to $M_{Pl}^2 = \frac{M_5^3}{k}\left(1-e^{-2\mu\pi}\right)  \simeq  M_5^3 / k$. 
The massive gravitons have a different normalization compared to the zero mode. This generates a different effective scale for these interaction which is given by  $\Lambda= M_5^{3/2} \sqrt{r_c}/\psi_n(\pi)$. In the large $\mu$ limit this reduces to $\Lambda = M_{Pl}\, e^{-\mu \pi}$ such that the interaction strength  with the massive gravitons is exponentially larger than with the massless one. 
In this limit, the radion contribution to the interaction Lagrangian is given by
\begin{equation}
    \mathcal{L}_{\text{int,r}}^{(1)}=\frac{1}{\sqrt{6}\Lambda}rT \ ,
\end{equation}
where $T=\eta^{\mu\nu}T_{\mu\nu}$ is the trace of the energy-momentum tensor of the matter field.

In addition, we need the three-point interactions between gravitons and radions and the next higher term in the expansion of the interaction with matter. The Lagrangian and the Feynman rules for these can be found in\cite{deGiorgi:2020qlg}. Here we only briefly comment on the coefficients of these interactions.
The couplings between KK-gravitons and the radion can be obtained by plugging into the Lagrangian the solutions $\psi_n$ and $\psi_r$ and integrating out the 5D.  In the case of interest only the 3-particles couplings are relevant, and in the limit $e^{-\mu\pi} \ll 1$ they reduce to  $\Lambda^{-1}$ times a numerical coefficient that depends on the type of particles involved. They are summarized in Fig. \ref{vertices--}.

\begin{center}
\begin{minipage}[h]{.1\textwidth}
 \begin{figure}[H]
      \includegraphics[scale=0.18]{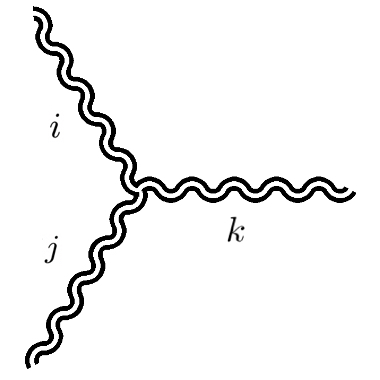}
\end{figure}
\end{minipage}
\begin{minipage}[h]{.1\textwidth}
\begin{flalign*}
   \quad\quad\propto \frac{\chi_{ijk}}{\Lambda} &&
    \end{flalign*}
\end{minipage}
\begin{minipage}[h]{.1\textwidth}
 \begin{figure}[H]
       \includegraphics[scale=0.18]{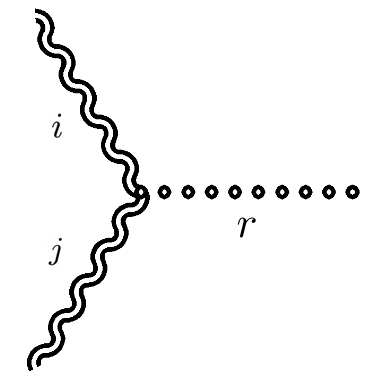}
\end{figure}
\end{minipage}
\begin{minipage}[h]{.1\textwidth}
\begin{flalign*}
   \quad\quad\propto \frac{\tilde{\chi}_{ijr}}{\Lambda}\left(\frac{m_i m_j}{\gamma_i \gamma_j}\right) &&
    \end{flalign*}
\end{minipage}
\begin{minipage}[h]{.1\textwidth}
 \begin{figure}[H]
       \includegraphics[scale=0.18]{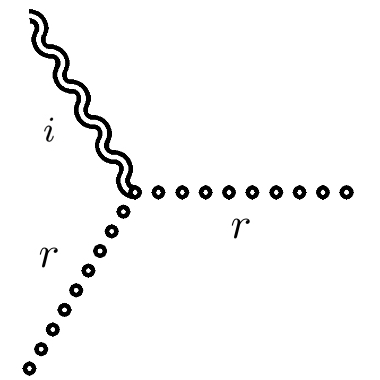}
\end{figure}
\end{minipage}
\begin{minipage}[h]{.24\textwidth}
\begin{flalign*}
   \quad\quad \propto \frac{\chi_{irr}}{\Lambda} &&
    \end{flalign*}
\end{minipage}
    \captionof{figure}{Cubic interactions between gravitons and radions.}
    \label{vertices--}
\end{center}
The numerical coefficients are given by
\begin{equation}
    \label{def:couplings}
    \begin{split}
        \chi_{ijk}&\equiv \frac{-2}{J_0(\gamma_i)J_0(\gamma_j)J_0(\gamma_k)}\int\limits_0^1 \diff{u} \ u^3 J_2(\gamma_iu)J_2(\gamma_ju)J_2(\gamma_ku)\ ,\\
         \tilde{\chi}_{ijr}&\equiv 2\frac{\gamma_i \gamma_j}{J_0(\gamma_i) J_0(\gamma_j)}\int\limits_0^1 \diff{u} \ u^{3}  J_1\left(\gamma_i u\right) J_1\left(\gamma_j u\right)\ ,\\
          \chi_{irr}&\equiv \frac{-2}{J_0(\gamma_i)}\int\limits_0^1 \diff{u} \ u^3 J_2(\gamma_iu)=\frac{8}{\gamma_i^2}\ .
    \end{split}
\end{equation}
The numerical coupling of the three radions vertex is trivially found to be $\chi_{rrr}=1/2$.

In the original RS model, the radion is massless which implies a long-range force much stronger than gravity. This is clearly in contradiction with observations, and, therefore, a realistic model has to include a radion mass. One popular possibility to generate as mass is the 
Goldberger-Wise mechanism \cite{Goldberger:1999uk} which relies on a new bulk scalar to stabilizes the distance between the branes. Alternative ideas for the stabilization of extra-dimensions based on other kinds of bulk matter have also been put forwards, see e.g. \cite{Luty:1999cz,Ponton:2001hq}. We remain agnostic as to the origin of the radion mass and introduce its mass by hand.

Finally, it should be mentioned that a singlet scalar could also interact with the SM via the Higgs portal operator
\begin{align}
\mathcal{L}_{\text{int}} \supset \lambda_{h\phi} H^\dagger H \phi^2\,.
\end{align}
This interaction is very well studied, see e.g. \cite{Cline:2013gha,Athron:2017kgt}, and we do not include it in our analysis since we want to keep our focus on the implications of higher dimensional gravity.

\end{section}

\begin{section}{Dark Matter Production}
\label{sec:DM_production}

Astrophysical and cosmological observations have clearly established the need for a large DM component in our Universe. The best measurement of its abundance comes from observations of the Cosmic Microwave Background (CMB) with the final analysis of the full  Planck data yielding $\Omega h^2 = 0.1200(12)$ \cite{Akrami:2018vks}. This result continues the best-known aspect of DM and, consequently, correctly reproducing it is one of the key marks of a successful theory. There are a number of mechanisms that explain how DM can be created in the early Universe.  
In the following, we consider the thermal freeze-out scenario while other possibilities such as a freeze-in will be discussed elsewhere. In this picture, DM is in thermal equilibrium with the SM plasma in the early Universe and decouples once the Hubble rate $H$ becomes comparable to the creation and destruction rates of the DM particles in the thermal environment.   
The evolution of the DM number density $n$ is described by the Boltzmann equation \cite{Kolb:1990vq}
    \begin{equation}
    \label{eq:BE}
        \Dot{n}+3Hn = -\left<\sigma v \right>\left( n^2 - n_{eq}^2 \right) \ ,
    \end{equation}
where $\left<\sigma v \right>$ is the thermally averaged cross section 
while $n_{eq}$ is the DM equilibrium density. For temperature $T$ small compared to the DM mass $m$, $\langle \sigma v \rangle$ is given by \cite{Gondolo:1990dk}
\begin{align}
    \left< \sigma v \right> &= \frac{1}{8m^4T \ K_2^2\left(m/T\right)}\int\limits_{s_{\text{min}}}^\infty \text{d}s \  \sigma(s) (s-4m^2)\sqrt{s}\ K_1\left(\sqrt{s}/T\right) \nonumber \\
    &= \frac{4x}{ \ K_2^2\left(x\right)}\int\limits_{1}^\infty \text{d}w \  \sigma(4m_\phi^2 w) (w-1)\sqrt{w}\ K_1\left(2x\sqrt{w}\right) \ ,
    \label{eq:thermal_average}
\end{align}
where $K_n(z)$ is the $n$-th Bessel-$K$ function. In the second line, we have rewritten the expression in terms of the dimensionless variables $w=s/4m^2$ and $x=m/T$ for convenience. Away from special kinematic configuration such as thresholds or resonances $\langle \sigma v \rangle$ can often be approximated by a polynomial in $v^2$ or, equivalently, $1/x$. 
Due to the KK-tower resonances and thresholds are ubiquitous in our model and we always use the full thermal average in our analysis. We also do not report approximate expressions in the velocity expansion but prefer to present $\sigma(s)$ instead.
Two classes of final states contribute to the $\phi$ annihilation rates, SM particles, and gravitational fields (i.e. gravitons and radions).  
\subsection{SM final states}
Annihilations into SM fields are mediated by the exchange of KK-gravitons and the radion in the s-channel \footnote{In principle also the massless graviton contributes but since its interaction strength is fixed by $M_{Pl}$ it can be neglected here.}. A set of representative Feynman diagrams is shown in Fig.\ref{SMAnnihilation}.
\begin{figure}[tb]
\centering
\includegraphics[width=0.7\textwidth]{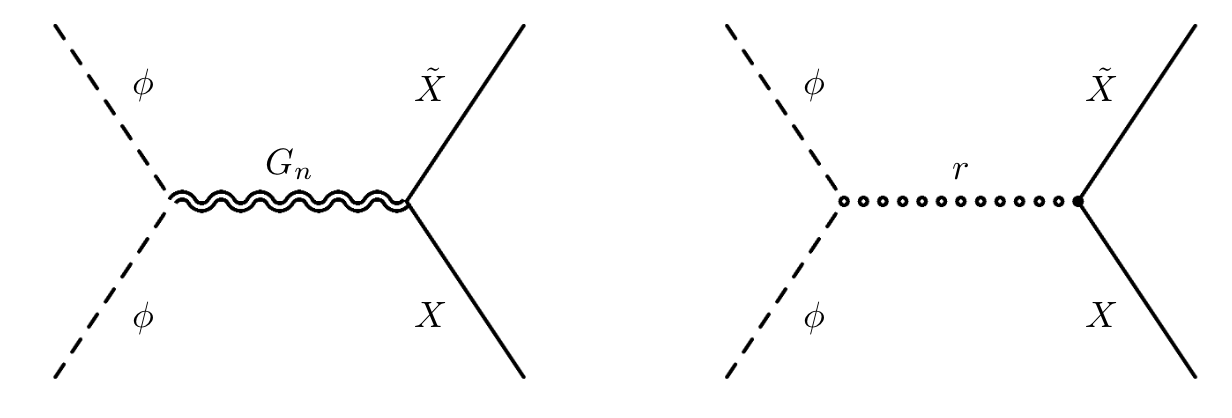}\par\vspace{0.1cm}
\caption{Annihilation channels of DM into SM particles via gravitons and radion exchange. Here $X$ and $\bar{X}$ should be read as a placeholders for all kinds of SM particles.}\label{SMAnnihilation}
\end{figure}
If the DM is heavy such that the masses of SM particles can be neglected one finds a relatively simple expression for the cross section
\begin{equation}
\label{cross:DMtoSM_grav}
\sigma(\phi\phi\overset{G_n}{\rightarrow}\text{SM SM})\approx\frac{73 }{1440 \pi   } \left(1-\frac{4 m_{\phi }^2}{s}\right)^{3/2}\frac{s^3}{\Lambda^4}\left|\sum\limits_n \frac{1}{s-m_n^2+i \Gamma_n m_n}\right|^2 \ ,
\end{equation}
where $\Gamma_n$ is the width of the graviton. In this limit a  good approximation of the width is $\Gamma_n \approx 0.0968 \frac{m_n^3}{\Lambda^2}$; more detailed expressions that take all masses into account can be found in Appx. \ref{app:decay}.
The whole KK-tower contributes and the cross section is characterized by a series of resonances. The thermal average smooths these out a bit but the resonance structure
remains clearly visible in $\langle \sigma v \rangle$. 
In addition, there is a contribution from radion exchange  which does not interfere with the gravitons. If the masses of the final state particles are neglected 
 one finds
\begin{equation}
\label{cross:DMtoSM_rad}
\sigma(\phi\phi\overset{r}{\rightarrow}\text{SM SM})\approx \frac{1 }{288 \pi }\left(1-\frac{4 m_{\phi }^2}{s}\right)^{-\frac{1}{2}} \left(1+\frac{2 m_{\phi }^2}{s}\right)^2 \frac{s^3}{\Lambda ^4}\left| \frac{1}{s-m_r^2+i \Gamma_r m_r}\right|^2 \ ,
\end{equation}
where we have neglected the contribution of massless gauge bosons. These do not couple to the radion at tree level and their contribution at freeze-out is small. 
Expressions for the cross section into different final states from both graviton and radion exchange with the full mass dependence can be found in \cite{Folgado:2019gie}.
For illustration, we show $\langle \sigma v \rangle$ from radion and KK-graviton exchange for a representative set of parameters and two radion masses in Fig.~\ref{fig:sigmav_rad_grav}.
\begin{figure}[t]
\centering
\includegraphics[width=.75\textwidth]{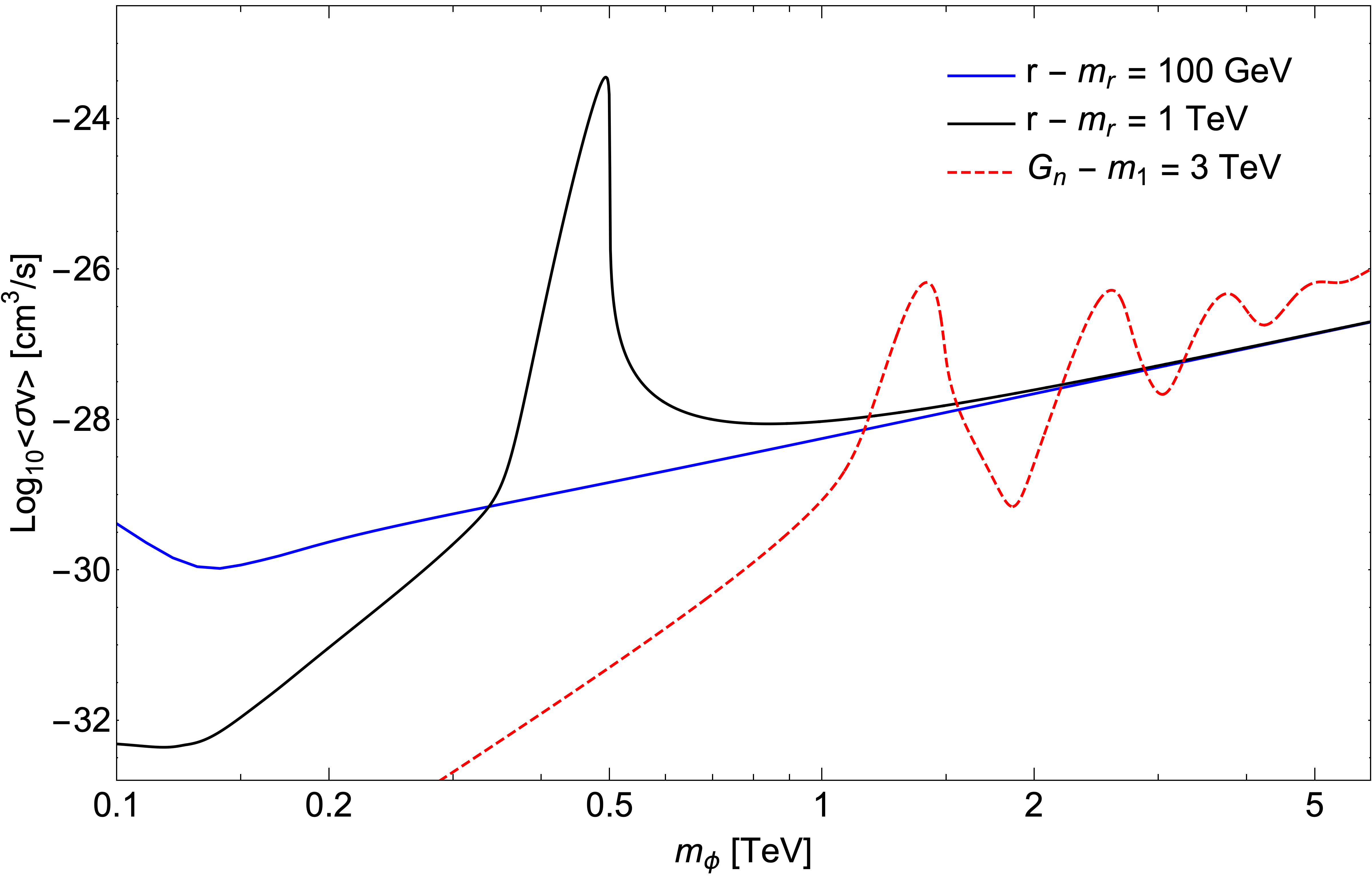}\par
\caption{Thermally averaged annihilation cross section into SM final states as a function of the DM mass. We choose $m_1=3$ TeV and $\Lambda = 8$ TeV. The contribution from KK-graviton exchange is shown in red (dashed). For the radion we show two representative choices of the radion mass, $m_r=100$ GeV and $m_r=1$ TeV in blue (solid) and black (solid), respectively. We fixed the temperature to $T=m_\phi/20$ which is in the ballpark of the typical freeze-out temperature.  \label{fig:sigmav_rad_grav}}
\end{figure}
The radion cross section away from the resonance is dominantly s-wave, and, therefore depends only mildly on temperature. In contrast, the graviton cross section is either dominated by the resonances or has the leading contribution at d-wave, i.e. $\langle \sigma v \rangle \propto T^2/m_\phi^2 $. Therefore, the temperature has a very strong impact on the thermal average in this case. For illustration, we use $T=m_\phi/20$ which is in the ballpark of the expected freeze-out temperature.   
As can be seen, the radion and graviton cross section are comparable and the relative importance is set by the position of the resonances peaks. The radion peak is more pronounced than the graviton peaks since the radion width is smaller but also the impact of the different gravitons can be seen in the form of distinct peaks.  
At $\sqrt{s}$ much bigger than $m_1$ these start to wash out slightly and the cross section approaches a continuum that is dominated by graviton exchange.

\subsection{Graviton and radion final states}
\begin{figure}[tb]
\centering
\includegraphics[width=0.95\textwidth]{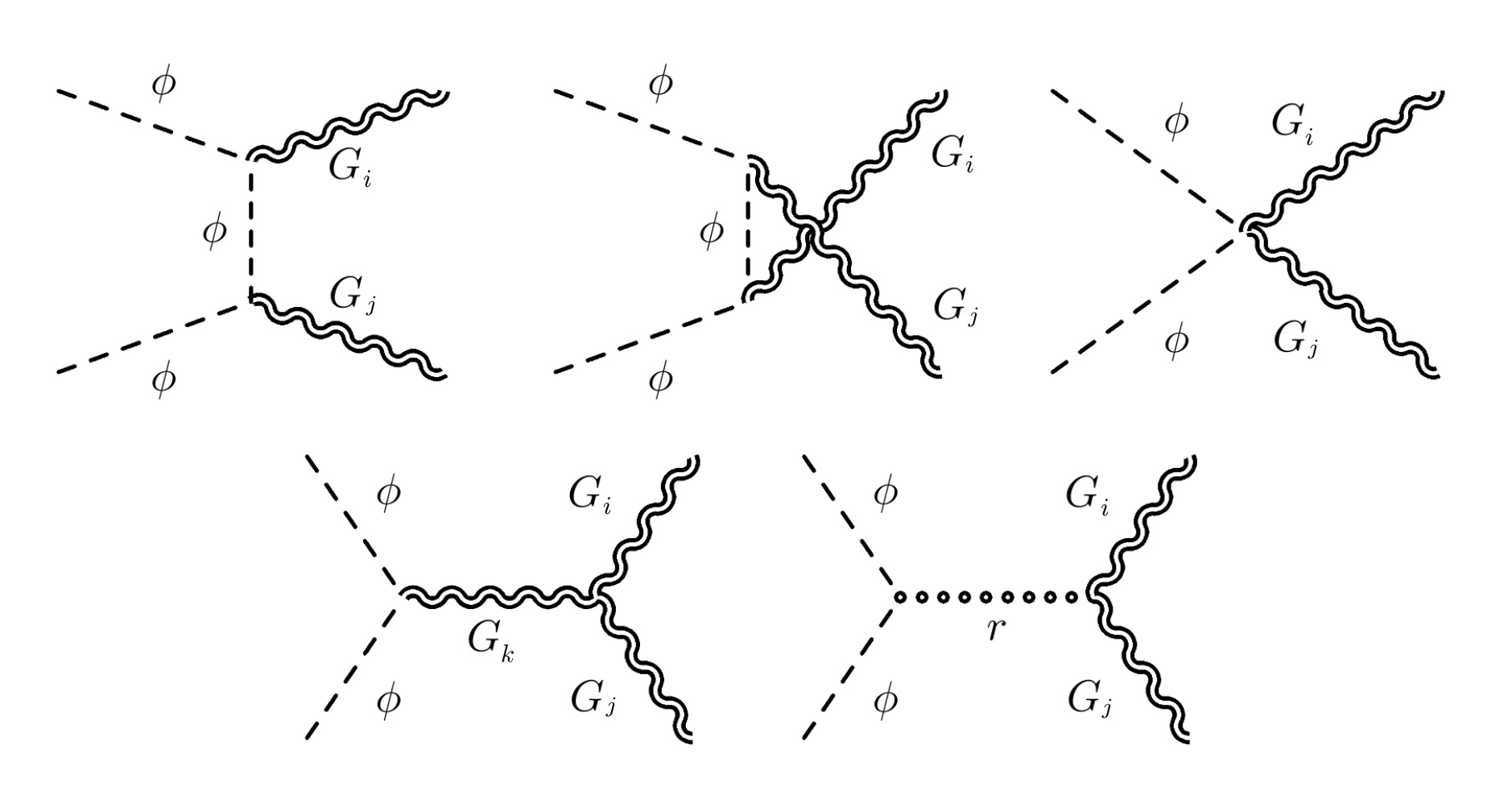}
\caption{Feynman diagrams contributing to $\phi\phi\longrightarrow G_iG_j$.}\label{GiGjprod}
\end{figure}
The situation is more complicated when gravitons and radion in the final states are considered. In contrast to annihilation into SM fields which only depend on the interaction between the gravitons and the energy-momentum tensor we now need to include also the self-interaction of the gravitons, the radion-graviton coupling, and the effective four-point interactions $\phi\phi G_i G_j$, $\phi\phi G_i r$ and $\phi\phi rr$ in our analysis. A set of representative Feynman diagrams for the case of annihilations into two gravitons is shown in Fig.\ref{GiGjprod}. The relevant Feynman rules and expressions for the coefficients of the graviton/radion interactions are reported in the appendix of \cite{deGiorgi:2020qlg}. 
Treating processes with external gravitons correctly is subtle. In the following, we briefly recapitulate the results of \cite{deGiorgi:2020qlg} where we discussed the challenges associated with graviton production in the Randall-Sundrum in detail, see also \cite{Bonifacio:2019ioc,Chivukula:2020hvi} for a related discussion on graviton scattering.

The longitudinal mode of the polarization tensor of massive spin-2 fields is proportional to $s/m^2$ which leads to a strong growth of the matrix elements with external gravitons in the high energy limit. 
Taking only the lightest graviton into account the matrix element $\phi\phi \rightarrow G_1 G_1$ grows as $\mathcal{M}\propto s^3$. Specializing to the case of DM annihilations this would imply a very strong growth of the associated cross section with $m_\phi$, see for example \cite{Lee:2014caa,Folgado:2019sgz}. This behavior is not altogether unexpected since 
an even worse growth of the amplitudes has been observed in spin-2 particle scattering amplitudes before \cite{Schwartz:2003vj,Christensen:2014wra,Bonifacio:2019mgk}.
In the absence of additional new physics, this leads to a breakdown of perturbative unitarity at a scale well below the naive effective scale of the theory $\Lambda$ \cite{ArkaniHamed:2002sp}. This is puzzling as the 5D graviton whose 4D decomposition leads to the Randall-Sundrum model does not have this issue.
The seeming contradiction here can be resolved by realizing that the 4D theories derived from higher dimensional models consist of a KK-tower of gravitons and not an isolated massive spin-2 field. Once, the full content of the theory is considered the contributions of the different gravitons to s-channel exchange conspire to cancel the contributions to the matrix elements that grow faster than $s$ \cite{Chivukula:2020hvi,Bonifacio:2019ioc,deGiorgi:2020qlg}. This is conceptually similar to the unitarization of massive vector boson scattering in the SM~\cite{Lee:1977eg} and has been dubbed unitarization by geometry in \cite{Bonifacio:2019ioc}. The chief difference with the usual Higgs mechanism is that a single field is not sufficient to restore perturbativity of the amplitudes in the EFT and the whole KK-tower is needed to fill the role of the SM Higgs~\footnote{ Additional complications arise if the scalar mode of the 5D graviton acquires a mass but it has recently been shown that these results hold if the popular Goldberg-Wise mechanism for generating the radion mass in the Randall-Sundrum model is treated carefully~\cite{SekharChivukula:2021btd}.}.

This has important implications for the annihilation rates into pairs of gravitons and graviton-radion final states. Unless the full tower of KK-gravitons is considered in the matrix elements unphysical contributions with a strong dependence on the center of mass energy arise. These can easily dominate the full cross section and lead to erroneous results. In practice, it is difficult to perform the sum over all internal KK-gravitons analytically. Therefore we use numerical methods when we want to include the full mass dependence. In the following, numerical results are calculated with a truncated KK-tower that includes the first 20 gravitons.  This suppresses the prefactors of the unphysical contributions substantially and ensures that their contribution to the annihilation rates can be neglected throughout the parameter space considered here.
Interestingly, in the high energy limit, analytic relations between the KK-graviton couplings derived in \cite{deGiorgi:2020qlg} can be applied at the cross section level. We find a simple expression for the cross-section for $G_n$-pair production in this limit
\begin{equation}
\label{eq:G1G1}
        \sigma\left(\phi\phi\longrightarrow G_n G_n \right) \approx \frac{\left(c_{n,4}^2-4 c_{n,4}+436\right)}{207360 \pi  }\frac{s}{\Lambda ^4}  = \frac{13}{5760\pi} \frac{ s}{\Lambda ^4}\ ,
\end{equation}
where the equality holds since the coefficients of the three-graviton vertex fulfill the relation (see Appx. \ref{app:sum})
\begin{align}
c_{n,4}\equiv \sum\limits_{m=1}^\infty \chi_{nnm} \left(\frac{\gamma_m}{\gamma_n}\right)^4=8\ .
\end{align}
 Thus, the high energy limit of $\sigma(\phi\phi\longrightarrow G_n G_n)$ is universal and does not depend on the produced graviton. 
Analogous calculations for the production of different gravitons lead to
\begin{equation}
    \sigma\left(\phi\phi\longrightarrow G_l G_k \right) \approx  \frac{13}{2880\pi} \frac{ s}{\Lambda ^4} \ .
\end{equation}
Similar considerations hold for the production of a radion and a KK-graviton. In the high energy limit the cross-section is found to be
\begin{equation}
   \sigma\left(\phi\phi\longrightarrow r G_n \right) \approx
 \frac{d_{n,0}^2}{2880 \pi}\frac{s}{\Lambda^4} =\frac{1}{2880 \pi}\frac{s}{\Lambda^4} \ ,
\end{equation}
where the equality holds since the radion-graviton-graviton coefficients involved fulfill the relation (see Appx. \ref{app:sum})
\begin{equation}
    d_{n,0}=\sum\limits_{m=1}^\infty \frac{\tilde{\chi}_{nmr}}{\gamma_n^2}=1 \ .
\end{equation}
Lastly, the cross-section for radion pair production is
\begin{equation}
   \sigma\left(\phi\phi\longrightarrow r r \right) \approx
 \frac{e_0^2}{5760 \pi}\frac{s}{\Lambda^4} =\frac{1}{5760 \pi}\frac{s}{\Lambda^4}\ ,
\end{equation}
where the equality holds due to the fact that (see Appx. \ref{app:sum})
\begin{equation}
    e_0= \sum\limits_{m=1}^\infty \chi_{mrr} = 1\ .
\end{equation}

In the regime that is most relevant for freeze-out, i.e. $\sqrt{s} \approx m_\phi$, we cannot use the above approximations directly. However, one can instead parametrize $s=4m_\phi^2 w$
and perform an expansion in $m_\phi$. Keeping only the leading contributions in the small parameter $w-1$ one arrives at 
\begin{align}
\label{eq:1-w_approx_1}
    \sigma\left(\phi\phi\longrightarrow G_n G_m \right) &\approx \frac{19 m_{\phi }^2}{576 \pi  \Lambda ^4 \sqrt{w-1}} \left[1-\frac{217}{114}(w-1)+\frac{7927 }{2280}(w-1)^2 \right] \\
    \sigma\left(\phi\phi\longrightarrow G_n G_n \right) &\approx \frac{19 m_{\phi }^2}{1152 \pi  \Lambda ^4 \sqrt{w-1}} \left[1-\frac{217}{114}(w-1)+\frac{7927 }{2280}(w-1)^2 \right] \\
    \sigma\left(\phi\phi\longrightarrow G_n r \right) &\approx \frac{m_\phi^2}{576 \pi  \Lambda ^4 \sqrt{w-1}} \left[1-\frac{19}{6}(w-1)+\frac{1453 }{120}(w-1)^2 \right]\\
    \sigma\left(\phi\phi\longrightarrow r r \right) &\approx  \frac{m_\phi^2}{1152 \pi  \Lambda ^4 \sqrt{w-1}} \left[1-\frac{19}{6}(w-1)+\frac{137 }{24}(w-1)^2 \right]
    \label{eq:1-w_approx_2}
\end{align}
Just as before the cross-sections are not sensitive to the type of KK-graviton(s) produced provided $m_\phi$ is significantly bigger them $m_n$. Furthermore, they are independent of the mass of the gravitons, i.e. of $m_1$. These simple expressions can be easily thermally averaged using Eq.~\ref{eq:thermal_average}. 
\begin{figure}[tb] 
\centering%
\subfigure[{}\label{fig:g1g1naive}]%
{\includegraphics[width=0.32\textwidth]{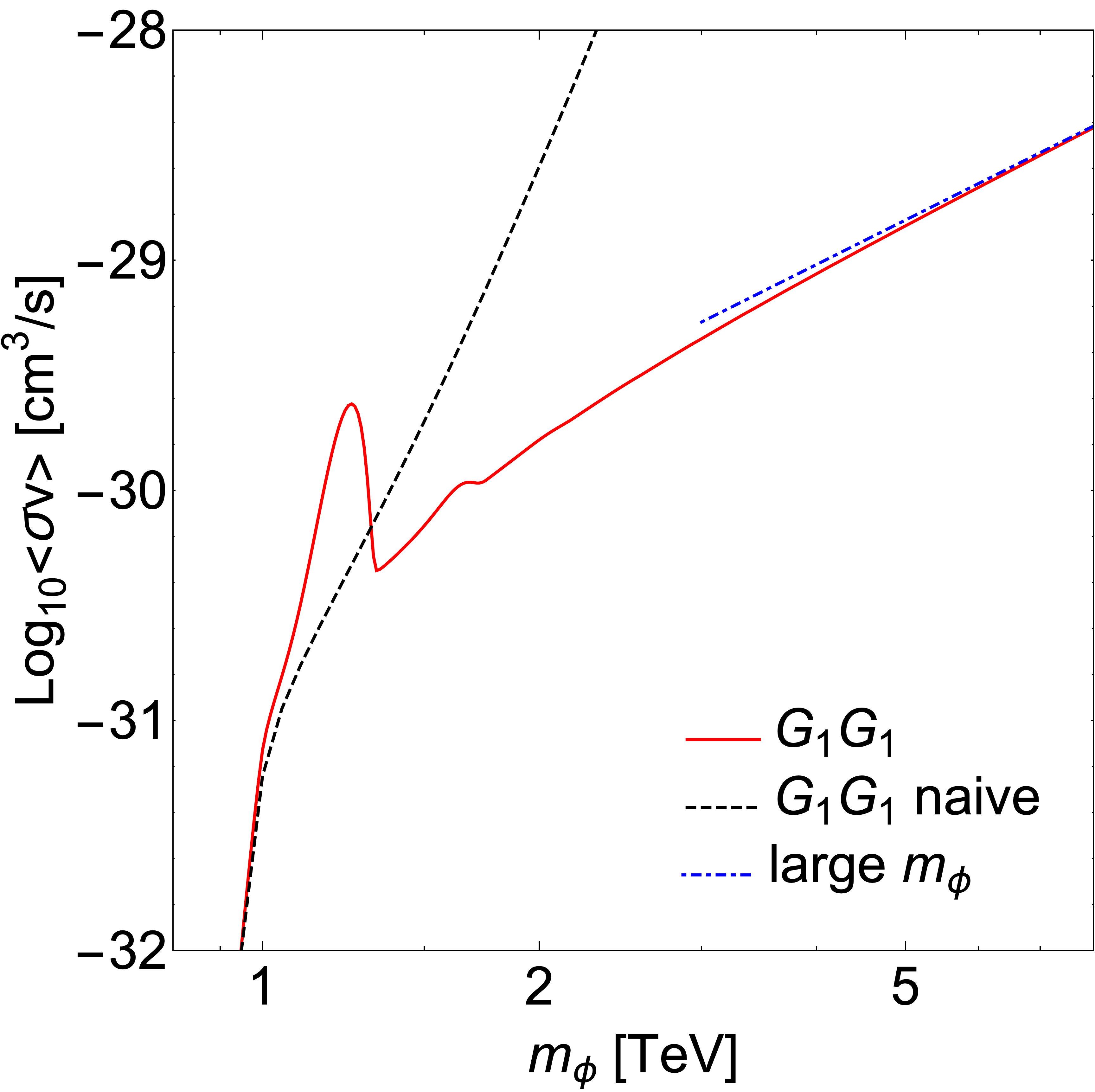}}\
\subfigure[{}\label{fig:g1g2naive}]%
{\includegraphics[width=0.32\textwidth]{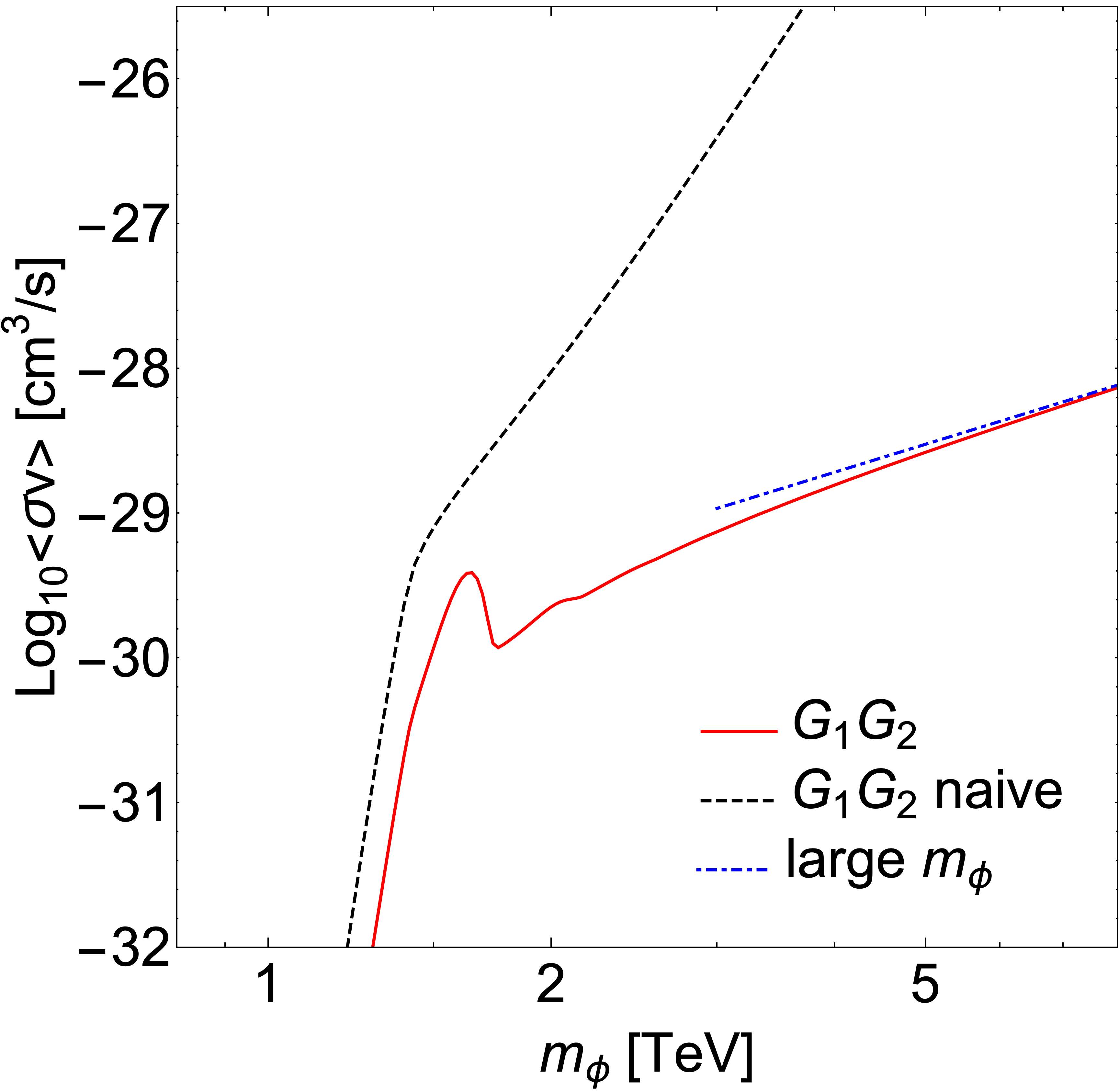}}\
\subfigure[{}\label{fig:g1rnaive}]%
{\includegraphics[width=0.32\textwidth]{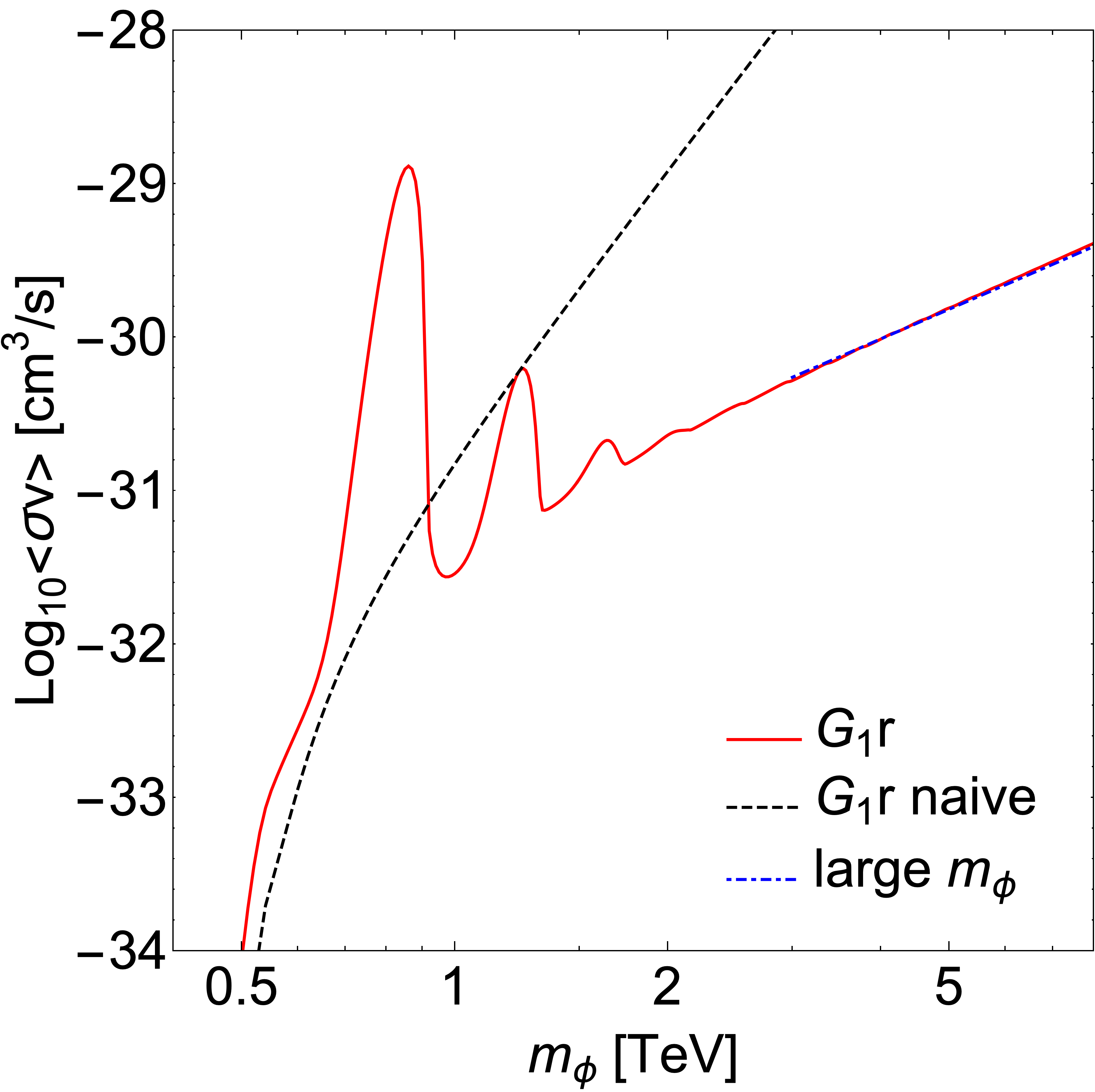}}
\caption{Thermally averaged cross section into the a) $G_1 G_1$, b) $G_1 G_2$, and c) $G_1$-radion final state for a representative set of parameters ($\Lambda=20$ TeV, $m_1=1$ TeV and $m_r=100$ GeV). The full numerical result obtained by including the KK-tower up to the first 20 gravitons is shown in red (solid) while the naive result obtained by truncating the tower after the highest graviton in the final states is shown in black (dashed). In blue (dot-dashed) we show our  approximation for the high $m_\phi$ limit. The temperature is taken to be $T=m_\phi/20$ as in Fig.~\ref{fig:sigmav_rad_grav}.}
\label{plotNaiveVSCool}
\end{figure}

For illustration we show $\langle \sigma v \rangle$ for an exemplary set of parameters in Fig.~\ref{plotNaiveVSCool}. The full numerical result is depicted by the solid red line. To put this into perspective we also include a naive approximation that truncates the KK-tower after the heaviest graviton in the final state (black dashed). As can be seen, the naive results grow very strongly with $m_\phi$ as expected from the high energy limit of the matrix elements discussed above\footnote{With $T_{fo}\approx m_\phi/20$ the average center of mass energy at freeze-out is approximately $m_\phi$.}. In contrast, the full result exhibits a much more modest $m_\phi$ dependence. For $m_\phi\approx 2 m_1$ the results already differ by an order of magnitude and, consequently,  computations that ignore the sum over the KK-tower can not be trusted almost immediately after the channel becomes kinematically accessible. In addition, we have also included the analytic large $m_\phi$, small $(w-1)$ approximation introduced above (blue, dashed). While this result is not able the capture to behavior for masses close to the threshold, the agreement with the numerical computation is quite good for $m_\phi\gtrsim 3 m_1$ and it reproduces the large mass behavior correctly.

\subsection{Total cross section}
\label{sec:total_xsec}
Let us now turn to the total annihilation cross section. The cross section into SM particles is well described by Eq.~\ref{cross:DMtoSM_grav} and  Eq.~\ref{cross:DMtoSM_rad} provided $m_\phi \gg m_t$. In the case of graviton and radion final states, the situation is a little bit more involved. Here, we have to take into account that more channels open as $m_\phi$ increases. The analytic approximations do not capture the channels with $m_n$ close to $m_\phi$. Nevertheless, if more channels are open the bulk of them can be captured by Eqs.~\ref{eq:1-w_approx_1}\--\ref{eq:1-w_approx_2}.
and the total cross section in RS-particles is given by
\begin{equation}
    \sigma_{RS}=\sum\limits_{i+j\leq2N}\sigma_{G_iG_j}+\sum\limits_{i\leq 2N} \sigma_{rG_i}+\sigma_{rr} \approx c N^2 \sigma_{G_1 G_1} \ ,
\end{equation}
where $G_N$ is taken to be the heaviest graviton pair that can be produced. Due to their larger multiplicity, the sums are dominated by the mixed $G_i G_j$ final. Identifying $N=\sqrt{s}/2\Delta m$ where $\Delta m$ is the mass spacing between the gravitons we find an estimate of total cross section into gravitational fields that captures the scaling in the large mass limit.
The parameter $c$ is an $\mathcal{O}(1)$ number that corrects for the fact that not all final state gravitons masses can be neglected in this sum. Weighting every channel by the phase space factor $\sqrt{\lambda(s,m_i^2,m_j^2)}/s$, where $\lambda$ is the K\"all\'en function, leads to the estimate $c\approx 0.6$ for large $N$. In the more relevant regime of $N\lesssim 20$  $c$ is slightly smaller and increases with $N$.  This approximation agrees well with our numerical results in the high mass regime.

In Fig.~\ref{plot:comparisonSV} we show a comparison of the thermally averaged annihilation cross section into SM and RS final states.  As can be seen the cross sections into SM particles dominates the total cross section for the temperature shown here. At low masses, the RS cross section is dominated by radion pair production before the first graviton radion and the graviton pair production take over. In this mass range, the final state with the largest accessible mass dominates but already at the third peak several processes have to be added and it is no longer possible to identify a single leading channel. 
Throughout the whole mass range the SM cross section exceeds the one into gravitons and radions. In the high mass regime, where the difference is smallest, $\langle \sigma v \rangle_{SM}\approx 3.5 \times \langle \sigma v \rangle_{RS} $ at $x=20$ which decreases to $\langle \sigma v \rangle_{SM}\approx 1.5 \times \langle \sigma v \rangle_{RS}$ at $x=30$. 
As these ratios are taken in the high-energy regime, they are independent of $m_1$ and $\Lambda$. Note, however, that this is close to the breakdown of perturbative unitarity.  Typically, freeze-out occurs at $x\approx25$.  It is thus reasonable to approximate the full cross section by the SM contribution as long as $m_1\gg m_t$ and a $\mathcal{O} (30\%)$ uncertainty in $\Omega h^2$ is acceptable.

\begin{figure}[tb]
\centering
\includegraphics[width=0.9\textwidth]{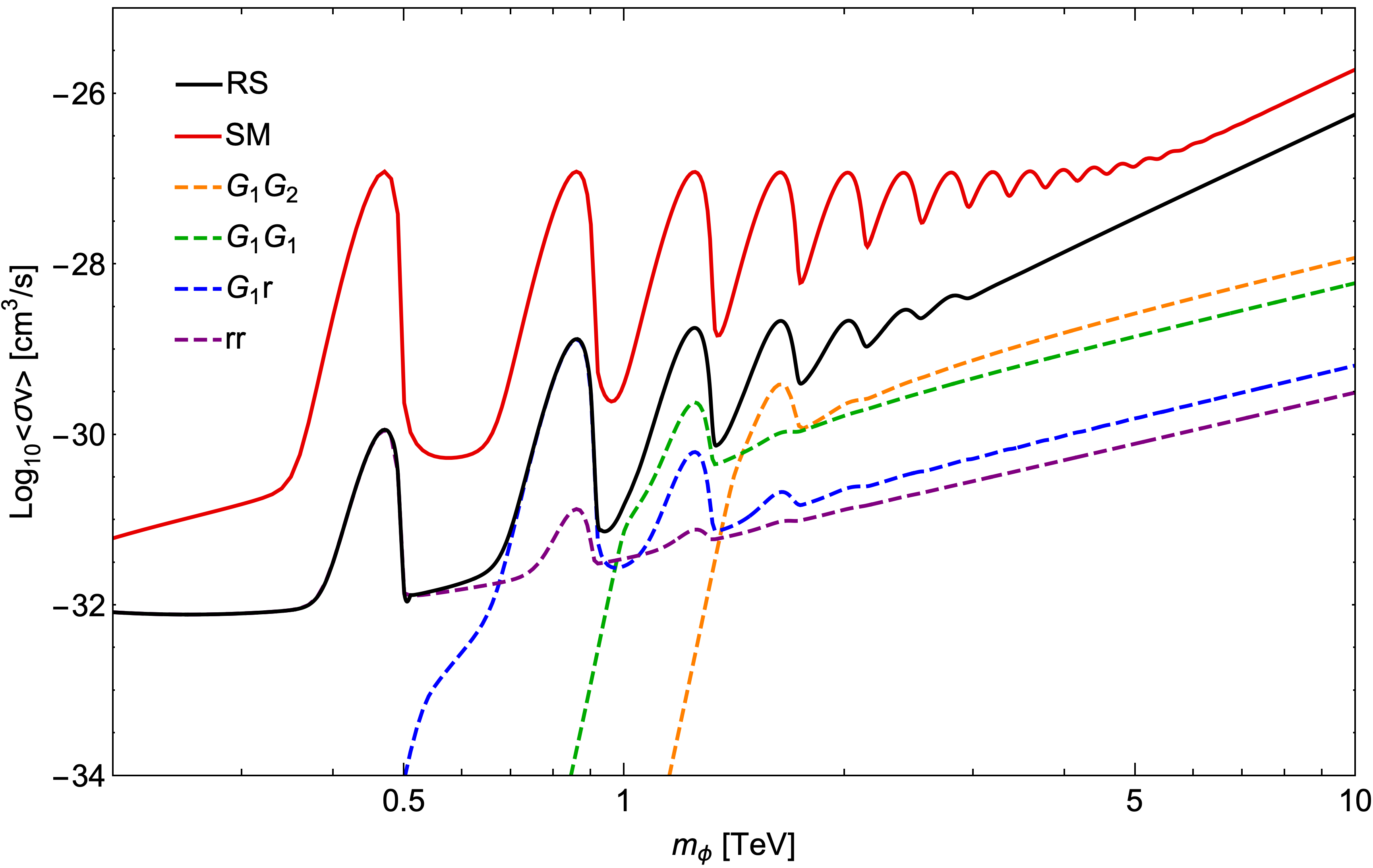}\par
\caption{Thermally averaged cross section into SM particles (red) and RS particles (black) final states for a representative set of parameters ($\Lambda=20$ TeV, $m_1=1$ TeV and $m_r=100$ GeV). The dashed lines correspond to the production of radion-radion (purple), $G_1$-radion (blue), $G_1 G_1$ (green) and $G_1 G_2$ (orange). The contribution from heavier gravitons is not shown separately but included in the black line. The temperature is taken to be $T=m_\phi/20$ as in Fig.~\ref{fig:sigmav_rad_grav}.
\label{plot:comparisonSV}} 
\end{figure}

\begin{section}{Phenomenology}

In this section, we first review some of the most stringent experimental and theoretical limits on the parameters of our model. Then we combine these with results for the cosmologically preferred parameter space of thermally produced DM and identify the regions that allow for a successful thermal freeze-out.

\begin{subsection}{Collider limits}
A stringent bound on the RS model comes from LHC experiments, in particular from searches for a heavy di-photon resonance.
We use the limits based on $37 \mbox{fb}^{-1}$ 
of $\sqrt{s}=13$ TeV data reported by ATLAS \cite{Aaboud:2017yyg}. In principle one could also consider the limits on the radion from searches for a spin-0 resonance. In practice, we find that these are not as strong and omit them from our analysis.

To extract limits on the parameters of the theory we compute the RS di-photon cross section with {\it CalcHEP} \cite{Belyaev:2012qa} and compare it to the limit presented by the collaboration. The experimental results are reported for some fixed value of $k/M_{Pl}$ which impacts the width of the resonance. However, throughout most of the parameter space the width of the graviton is small compared to the energy resolution and in the region where this is no longer the case the limit is insensitive to the width. Therefore, we can use the limits of \cite{Aaboud:2017yyg} directly.

\begin{figure}[tb]
\centering
\includegraphics[width=0.9\textwidth]{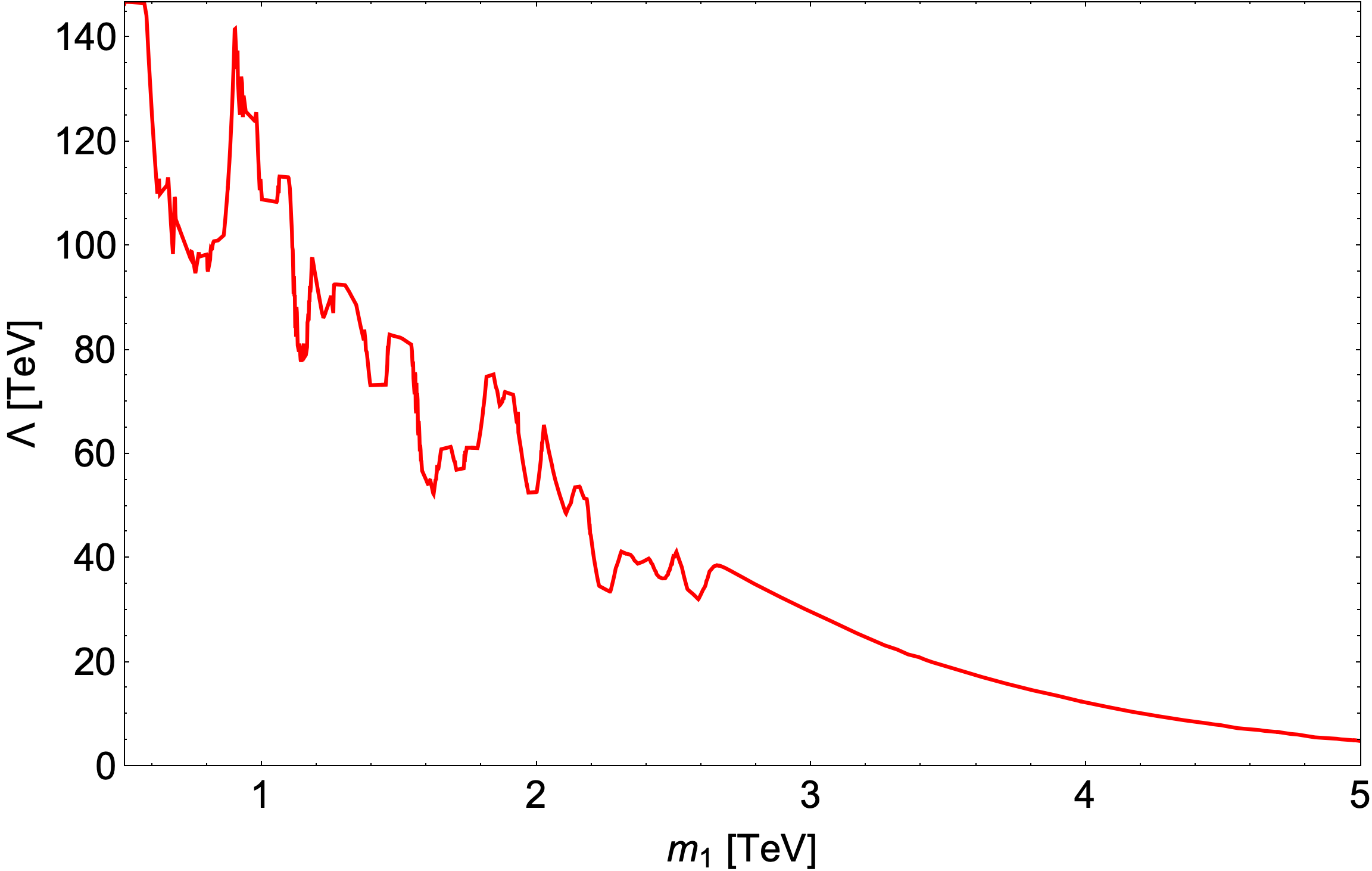}\par
\caption{Limit on $\Lambda$ as a function of $m_1$ inferred from the ATLAS di-photon resonance search.  }\label{cplotCH}.
\end{figure}
For illustration we show the limit on $\Lambda$ as a function of $m_1$ in Fig. \ref{cplotCH}. As can be seen, the bound is particularly stringent at low graviton masses and exceeds $100$ TeV for $m_1< 1$ TeV. It stays above $50$ TeV up to $m_1 \approx 2$ TeV. Once the mass of the graviton exits the energy window accessible for a resonance search, i.e. for $m_1 \geq 2.5$ TeV, the limit softens considerable. For $m_1\geq 5$ TeV the collaboration does not report any limits in the Randall-Sundrum model. In principle, one could attempt to extend them by recasting a limit on virtual graviton exchange in large extra-dimensions that comes from a different analysis of a subset of the full data. We do not attempt this here since we will find that the part of the parameter space where this becomes relevant is disfavored by theoretical considerations.      
\end{subsection}

\begin{subsection}{Perturbative unitarity}

Since we are working in an effective theory it is of great importance to understand where our calculations are valid.  The proportionality of the vertices' couplings between matter and gravitons (and between gravitons and radions) to $\Lambda^{-1}$ suggests that perturbativity ought to break at this scale. However, the KK decomposition obscures the fact that the 5D theory from which our model arises is already an effective theory with a breakdown scale that is set by 5D gravity with $M_5$ and the warp parameter $k$ as dimensionful parameters. Therefore, one might wonder which scale actually controls the validity of the 4D theory.
One possibility for assessing this question is perturbative unitarity.

We use partial wave perturbativity following the approach of e.g. \cite{Chanowitz:1978uj,Lee:1977eg,Kahlhoefer:2015bea}. The  helicity matrix element for the $J$-th partial wave with initial and final states $i$ and $f$ is given by
\begin{equation}
    \mathcal{M}_{if}^J(s) =\frac{1}{32\pi} \beta_{if} \int\limits_{-1}^1 \diff \cos{\theta} \  d_{\mu\mu'}^J(\theta) \mathcal{M}_{if}(s,\cos{\theta}) \ ,
\end{equation}
 where $\beta_{if}$ denotes a kinematical factor, $\theta$ is the scattering angle while $d_{\mu\mu'}^J(\theta)$ is the $J$-th Wigner-functions for total spins of initial and final states $\mu$ and $\mu'$.
Unitarity of the $S$-matrix enforces an inequality on the elastic elements 
\begin{equation}
\label{conditions}
    0 \leq \Im{\mathcal{M}_{ii}^J} \leq 1 \quad , \quad \Re{\mathcal{M}_{ii}^J} \leq \frac{1}{2} \quad \forall J\in \mathbb{N} \ ,
\end{equation}
which can be turned in an equality by diagonalizing the matrix $M_{ij}^J$. Therefore, the limit inferred from a single elastic scattering channel is conservative and a more stringent bound can be obtained by considering the full set of accessible states. 

For simplicity we 
 focus on the $J=0$ amplitude with $\mu= \mu' = 0$
and start with the elastic scattering of the scalar field $\phi$, see
Fig. \ref{elastic} for a representative set of the Feynman diagrams.
\begin{figure}[tb]
\centering
\includegraphics[width=0.8\textwidth]{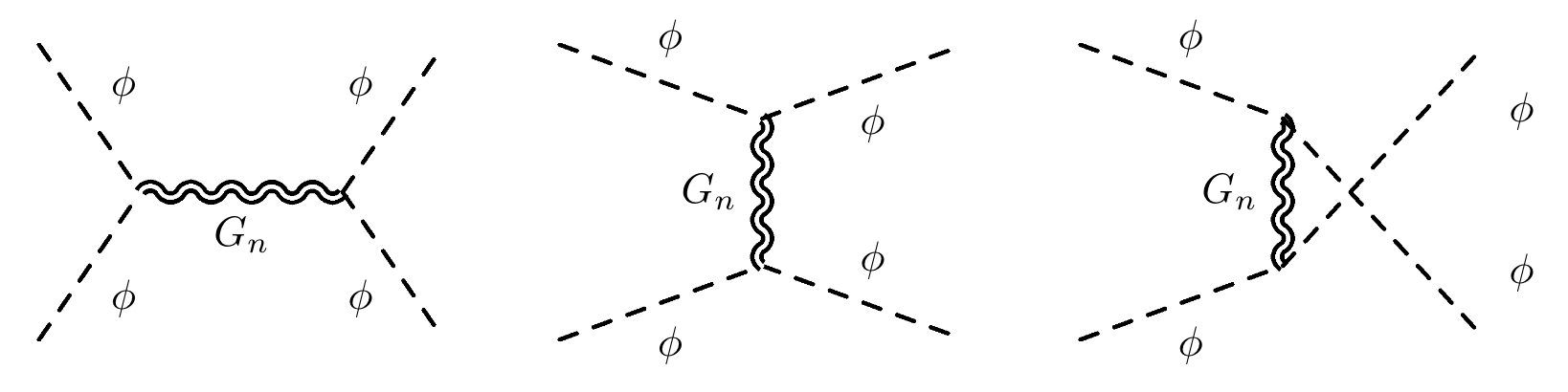}\par
\caption{Diagrams contributing to elastic scattering of scalar fields through $G_n$, $\phi\phi \overset{G_n}{\longrightarrow} \phi\phi$.}\label{elastic}
\end{figure}
For the sake of simplicity, we focus on the s-channel diagram for now \footnote{The other diagrams will be included analogously later on and do not change the qualitative result.}. The contribution from the full KK-tower to this amplitude is given by
\begin{equation}
    \mathcal{M}=\tilde{\mathcal{M}}\frac{1}{\Lambda^2}\sum\limits_{n=1}^\infty \frac{1}{s-m_n^2 + i \ m_n\Gamma_n} \equiv \tilde{\mathcal{M}}\mathcal{S}(s)\ .
    \end{equation}
Naively, one could try to estimate the sum by
assuming that the first $N$-gravitons have negligible mass compared to $\sqrt{s}$, i.e. $m_n \ll \sqrt{s}$. Taking $\Gamma_n \ll m_n$ and neglecting the contribution from heavier gravitons suggest the following scaling
\begin{equation}
\label{less-rough}
    \Lambda^2 \mathcal{S}(s)=\sum\limits_{n=1}^N \frac{1}{s-m_n^2 + i \ m_n\Gamma_n} \approx \sum\limits_{n=1}^N \frac{1}{s}=\frac{N}{s} \ . 
\end{equation}
The number $N$ is approximately given by the total energy divided by the mass-gap between each graviton $\Delta m$, which tends to a constant as the number of the related gravitons grow, i.e.
\begin{equation}
\label{rough}
    N \approx \frac{\sqrt{s}}{2\Delta m} \ .
\end{equation}
It follows that the sum of the propagators does not contribute as $1/s$ in the high energy regime, but rather as $N/s \approx 1/\sqrt{s}$, which makes the amplitude grow $\propto s^{3/2}$. A more rigorous way to arrive at this conclusion is to consider the large $s$   approximation of $\mathcal{S}(s)$ \cite{Kisselev:2005bv,Giudice:2004mg}
\begin{equation}
\label{ref:propsum}
    \mathcal{S}(s)\approx -\frac{\gamma_1 } {2 \Lambda ^2 m_1 \sqrt{s}} =-\frac{1 } {2 \bar{M}_5^3 \sqrt{s}} \quad \text{with} \quad \bar{M}_5\equiv M_5 e^{-\mu\pi} \ ,
\end{equation} 
which confirms our rough estimate. In contrast to the hand-waving argument above, the validity of this approximation is easily quantified and found to agree to better than $5\%$  for
\begin{equation}
\label{ref:better}
    \sqrt{s} \ \gtrsim \ 2.3 (\Lambda^2 m_1)^{1/3} \approx 3.5 \bar{M}_5\,.
\end{equation}
This implies a scaling $\mathcal{M} \propto s^{3/2}$  which points towards a breakdown of unitarity at a scale below $\Lambda$. Using the approximation  eq.~\ref{ref:propsum} and including the $t$- and $u$-channel contribution  
yields a perturbativity limit of
\begin{equation}
    \sqrt{s} \leq \left(\frac{160}{7} \pi\right)^{1/3} \bar{M}_5 \approx 4 \bar{M}_5 \,.
\end{equation}
Thus, the breakdown scale is set by the warped 5D Planck mass $\bar{M}_5$. A similar argument also holds for DM annihilations into SM particles. In this case the growth of $\mathcal{M}$ is also reflected in the large $\sqrt{s}$ behavior of the $\sigma_{SM} \propto s^2$, or, equivalently $\left<\sigma v\right>_{SM} \propto m_\phi^4$, that can be see in the upper right corner of Fig.~\ref{plot:comparisonSV}.

Even though it is intuitive that the scale up to which the 4D theory remains valid is set by the scale in the underlying 5D theory, it is slightly disconcerting that the result relies on the high energy limit of the sum of graviton propagators that only becomes a good approximation at $\sqrt{s}$ barely in the perturbative regime. Therefore, we aim to strengthen this result by considering a different part of the $S$-matrix, namely scattering of KK-gravitons. In the case of $G_i G_i \rightarrow G_i G_i$ most of the work has already been done by \cite{Chivukula:2020hvi}.
As mentioned above the leading contribution to this matrix element is proportional to $s^1$ which might cast some mild doubt on the $s^{3/2}$ scaling found in $\phi$ scattering. However, one has to take into account that increasing $\sqrt{s}$ also opens the thresholds for the production of more graviton pairs. Thus we cannot neglect inelastic scattering and we perform a coupled channel analysis following the approach of \cite{Chanowitz:1978uj}. 
\begin{figure}[tb]
\centering
\includegraphics[width=0.8\textwidth]{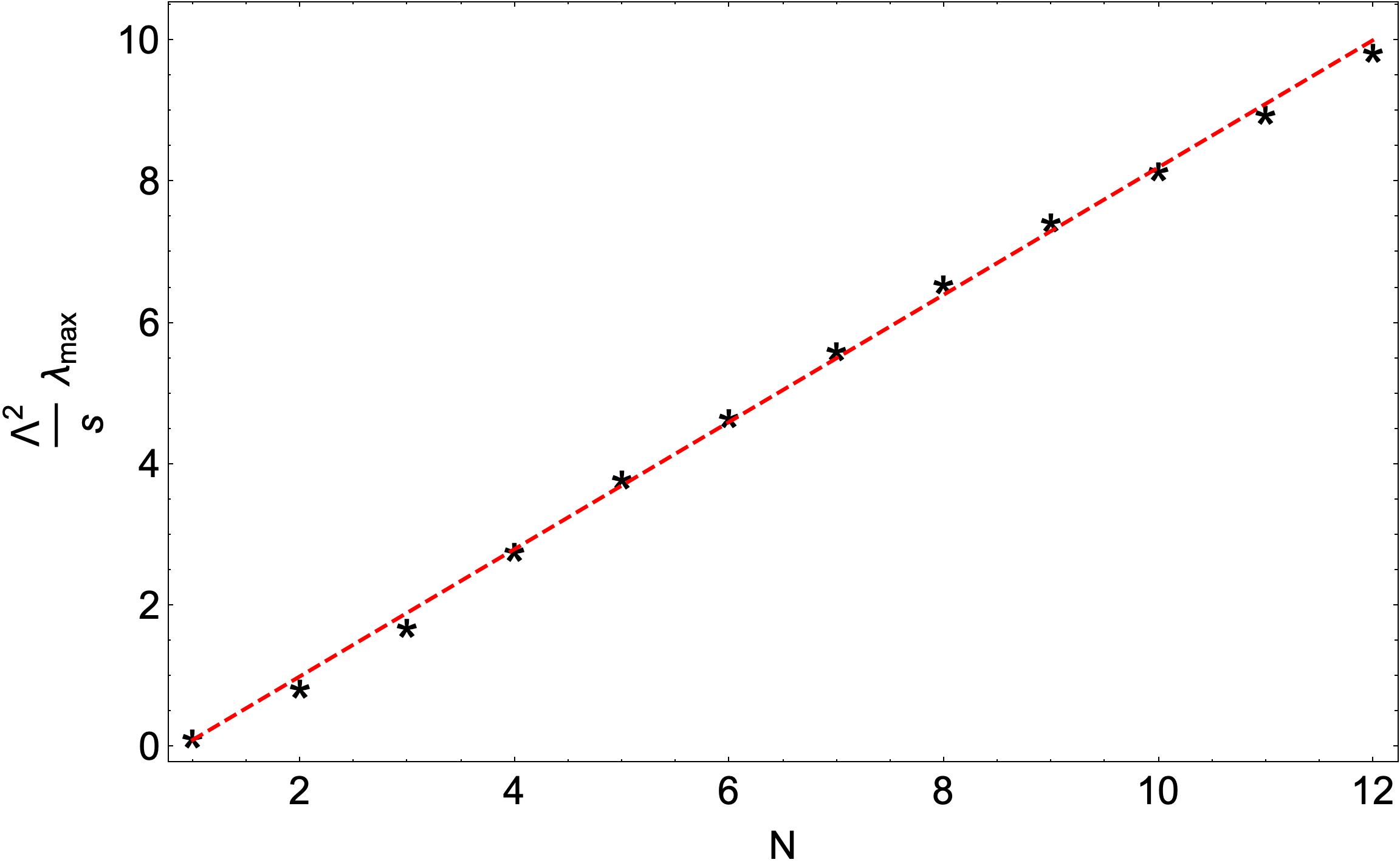}\par
\caption{Largest eigenvalue of the graviton scattering matrix $\lambda_{max}$ rescaled by $\Lambda^2/s$ as a function of $N$. The red-dashed line shows the linear fit that was used to extract the perturbativity limit.}\label{max-eigenvalues}
\end{figure}
We combine the results given in \cite{Chivukula:2020hvi} with our own computation to derive the  inelastic scattering  matrix elements for $G_i G_j \rightarrow G_k G_l$. As noted previously, these matrix elements grow faster than $s$ unless the appropriated sum rules for the coefficients of the three- and four-graviton are applied. Unfortunately, we have not found a simple analytic expression for these so we opt for a numerical analysis instead.
We confirm that the scaling of the amplitudes is reduced to $\mathcal{O}(s)$ if a  sufficient number of exchanged-gravitons are included. For computational reasons and since it suffices to our purpose, we consider only the 0-helicities of the gravitons and construct a matrix with all states  whose production is allowed at $\sqrt{s}$. Numerical evaluation of the partial wave amplitudes  then leads to the following scaling for the largest eigenvalue 
\begin{equation}
    |\lambda_{\text{max}}(N)| 
    \approx 0.9 N\frac{s}{\Lambda^2}
\end{equation}
for $\sqrt{s}\gg m_1$, where $N$ is the number of the heaviest graviton that can be pair produced. An illustration of this behavior together with our numerical results  as can be seen in Fig.~\ref{max-eigenvalues}.
Recalling that $N\approx \sqrt{s}/2\Delta m$ and extracting the proportionality constant from the numerical results we find the perturbativity condition  to be 
\begin{equation}
    \sqrt{s}\lesssim 1.9 \bar{M}_5 \,.
\end{equation}
This confirms and improves the previous result and, more importantly, is independent of the high energy limit of the sum of graviton propagators. 
\end{subsection}

\begin{subsection}{Parameter space favored by freeze-out}
\label{sec:parameter_space}
We focus our analysis on DM and lightest KK-graviton  heavier than SM particles since this region potentially allows for the production of DM via the freeze-out mechanism. It would be interesting to investigate to which extent a lighter and more weakly coupled lightest graviton is in agreement with experiments but since this is not expected to be consistent with thermally produced DM we do not consider this case further.
The annihilation cross section in this regime is dominated by SM final states, see Sec. \ref{sec:total_xsec}. 
For the efficient computation of the relic density, we implement the annihilations into SM particles via the first six gravitons in  \texttt{ MicrOMEGAs} \cite{Belanger:2013oya,Belanger:2006is}  and remove the artificially large contribution from RS final states by hand\footnote{Solving the Boltzmann equation with a numerical determination of the full annihilation cross section into graviton final states via summation of many gravitons exceeded our computational resources.}. We scan over  $m_\phi$
 and $m_1$ while keeping the radion mass fixed to $1$ TeV and extract the value of $\Lambda$ required for the thermal production of the observed relic density.

\begin{figure}[tb]
\centering
\includegraphics[width=0.70\textwidth]{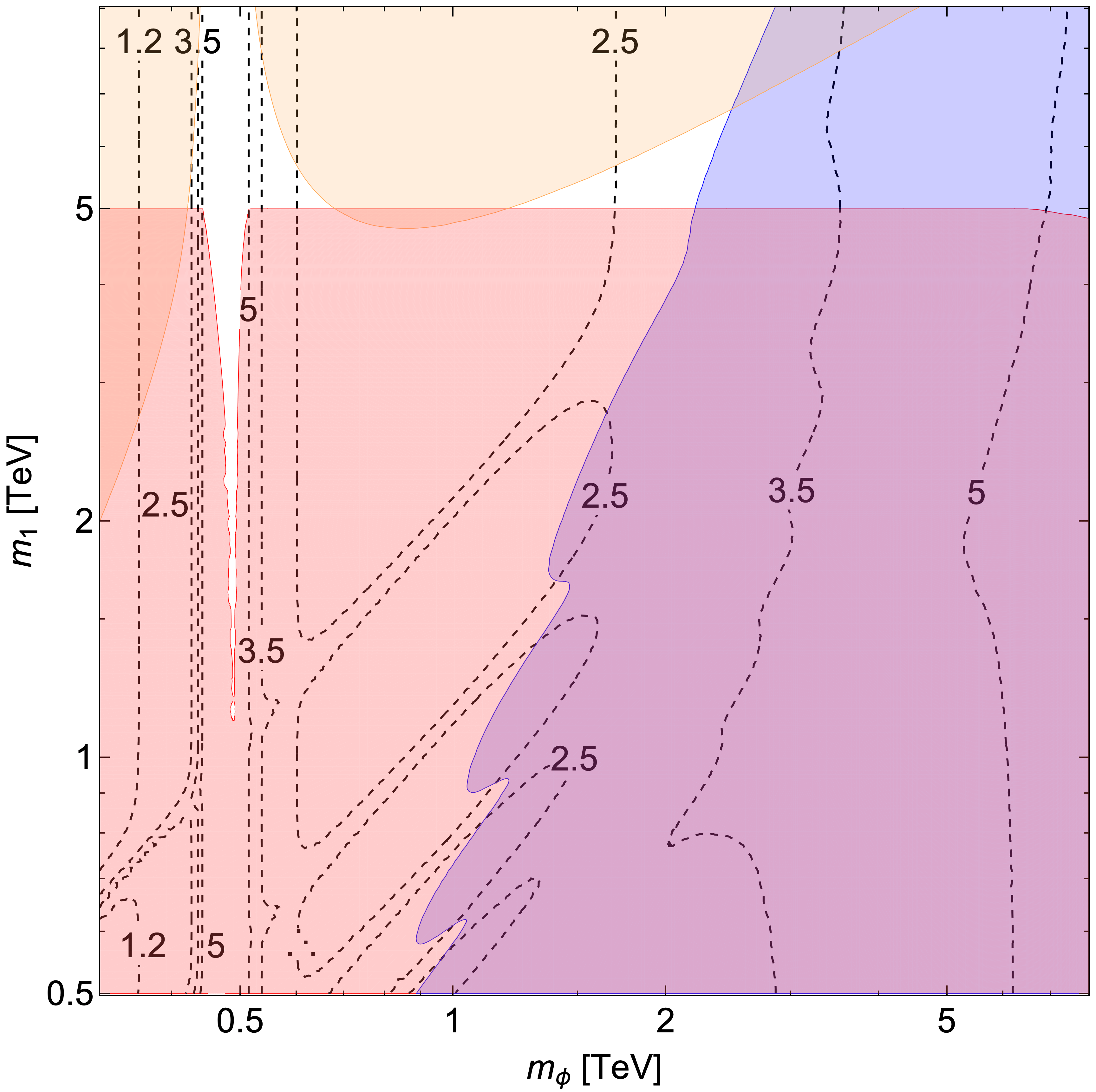}\par
\caption{
Experimental and theoretical limits on thermally produced real scalar DM in the RS model.
The values of $\Lambda$ [TeV] needed for $m_r = 1$ TeV to obtain $\Omega h^2_{DM}=0.12$ through freeze-out are indicated by the labels on the dashed contours. The red regions is excluded by LHC searches for di-photon resonances. The blue (orange) region is outside of the range of validity of perturbative computations due to breakdown of partial wave unitarity (large graviton width, $\gamma_1 \geq m_1/2$). In the white region thermal production of the observed relic density is viable. }\label{fig:plot_rad_grav}
\end{figure}

The results are visualized in Fig. \ref{fig:plot_rad_grav}. The dashed contours indicate lines of constant $\Lambda$. On top of this, we superimpose the limits from collider searches and theoretical consistency conditions. In the blue region starting at $m_\phi \gtrsim 2$ TeV, the value of $\Lambda$ preferred by freeze-out is in the non-perturbative region and hence our computation is not reliable. On the other hand, the strong collider bounds of Fig. \ref{cplotCH} excludes the thermal DM region with $m_1 \leq 5$ TeV almost completely. The only exception is for $m_\phi \approx m_r/2$ where the radion resonance allows for efficient annihilations even for comparatively large values of $\Lambda$.  In the top half, the majority of the parameter space is characterized by $\Gamma_1 \geq m_1/2$, which also casts severe doubts on the reliability of a perturbative computation.
This leaves two regions where thermal production remains viable, i.e. close to the radion resonance, and for $m_1$ slightly above 5 TeV. This second  small corner is uncomfortably close to the theoretical consistency conditions. While we can not exclude that a more complicated underlying theoretical model justifies $2 m_\phi = m_r$ or that the perturbativity limits are close to being violated in nature, this makes it unlikely that thermal production in the minimal Randal-Sundrum is the origin of DM. Naturally, this statement cannot be generalized to extra-dimensional theories as a whole and more complex extra-dimensional constructions with fields beyond the graviton in the bulk remain viable, see for example \cite{Rueter:2017nbk,Carmona:2020uqx}.

\end{subsection}

\end{section}

\begin{section}{Conclusions}
\label{sec:Conclusions}

There is a great interest in new ideas for the interactions between DM and the SM. In this paper, we studied  DM interacting with the SM via a massive spin-2 mediator. 
In this context, an interesting complication arises as soon as the energy is sufficient to produce the mediators on-shell. The longitudinal modes of the polarization tensor of massive spin-2 fields are proportional to $E^2/m^2$ which leads to the rapid growth of their production rate with energy almost instantly above the threshold.
A similar behavior is known from studies of spin-2 scattering in massive gravity and is a sign of an early breakdown of perturbative unitarity in theories with a single massive tensor field.
Due to the well-known issues of theories with spin-2 fields, all studies of these interactions have to resort to an EFT approach. Here, we focused on an EFT that is generated by the compactification of a warped extra-dimensional model. Interestingly, such a model is expected to remain perturbative up to the scale of 5D gravity. Relying on our earlier work on the unitarization of spin-2 production amplitude in warped extra-dimensions \cite{deGiorgi:2020qlg}, we are able to show that the strong growth of the cross section above the threshold is not physical and gets reduced considerably once the full KK-tower is included in the computation. Thus, an EFT with a limited number of degrees of freedom is not able to capture the behavior of extra-dimensional theories, and the results obtained with this approximation are misleading. 
We derived the DM annihilation cross sections carefully taking the KK-tower into account and evaluated it numerically. To better analyze the high mass behavior and to cross-check our numerical study we also report approximate analytic expressions in this limit.  



We have used our results for the annihilations to reassess the parameter space that allows for a thermal production of the observed relic abundance. After confronting these results with limits from collider searches for the lightest KK-graviton and theoretical consistency conditions, we find that gravitationally interacting scalar DM is under more pressure than previously anticipated. The only region which is clearly viable features resonant annihilations via the radion and is therefore characterized by an unexpected tuning between the radion and the DM mass. In addition, we find a second small region that is dangerously close to both the experimental limit and the theoretical consistency conditions.
Given that a perturbative computation will likely be modified even before the breakdown of unitarity, it is difficult to make a definitive statement about this region without specifying the UV completion of extra-dimensional gravity. In any case, the upcoming LHC runs will improve the experimental limits and close the gap.

In this work, we mainly focused on the non-relativistic limit of the annihilation cross section which is relevant for thermal freeze-out. However, it is clear that the unitarization of the annihilation cross section by the KK-tower will also affect the cross sections in the regime that matters for freeze-in. A study of the impact of this production mechanism would therefore be very interesting.
Finally, we also want to note that there is an increasing interest in theories with DM of spin-2 or higher \cite{Criado:2020jkp,Falkowski:2020fsu}. While our analysis does not directly address these ideas one might expect that variations of the unitarity issues found here also persist there. Thus, the results derived with this approach could be more sensitive to unspecified UV physics than naively expected.

\end{section}

\section*{Acknowledgments} 

We thank the Max-Planck Institute of Physics (MPP) for access to some of the computational resources employed in the work.  

\end{section}

\begin{appendices}
\numberwithin{equation}{section}


\begin{section}{Decay widths}
\label{app:decay}
In this appendix, we present the decay widths of the gravitons and the radion. In both cases the decay into SM-particles and into other RS-particles in the final state are considered. 
\subsection{Graviton width}
The graviton decays into all SM particles, into lighter gravitons, and radions. Our results for the partial width into SM particles match those reported in \cite{Folgado:2019sgz}. The width of a heavy KK-graviton decaying to a light KK-graviton in a different geometry has been reported in \cite{Giudice:2017fmj}. Even though their result ought to match ours once the coefficients of the KK-graviton vertex are adapted to our geometry, we do not find agreement. The difference has been traced to the relative sign of the part $R_4(\tilde{g})$ and the contribution with a 5D derivative to the three-graviton vertex. We have checked our result by comparing it with the three-graviton vertex reported by \cite{Bonifacio:2019ioc}. In addition, we verified that the unitarization of the graviton production amplitude does not go through with the other sign.

The partial width for decays into a pair of Higgs bosons reads
\begin{equation}
    \begin{aligned}
         &\Gamma_{G_n  \rightarrow  H  H}= \frac{\sqrt{1-\frac{4 M_H^2}{m_n^2}} \left(m_n^2-4 M_H^2\right){}^2}{960 \pi  \Lambda ^2 m_n} \ .
    \end{aligned}
\end{equation}
For a Dirac fermion pair we find
\begin{equation}
    \begin{aligned}
        &\Gamma_{G_n  \rightarrow   \Bar{\psi}\psi }= N_c\frac{m_n^3\left(1-\frac{4 m_{\psi }^2}{m_n^2}\right)^{\frac{3}{2}} \left(1+\frac{8 m_\psi^2}{3m_n^2}\right)}{160 \pi  \Lambda ^2 }\ .
    \end{aligned}
\end{equation}
where $N_c$ is the number of colors.\\
For the massive gauge bosons, $Z$ and$W^\pm$ the width is
\begin{equation}
    \begin{aligned}
         &\Gamma_{G_n  \rightarrow  Z Z}=\frac{\sqrt{1-\frac{4 M_Z^2}{m_n^2}} \left(56 m_n^2 M_Z^2+13 m_n^4+48 M_Z^4\right)}{960 \pi  \Lambda ^2 m_n} \ ,\\
         \left.\right.\\
          &\Gamma_{G_n  \rightarrow  W^+ W^-}=\frac{\sqrt{1-\frac{4 M_W^2}{m_n^2}} \left(56 m_n^2 M_W^2+13 m_n^4+48 M_W^4\right)}{480 \pi  \Lambda ^2 m_n} \ ,
    \end{aligned}
\end{equation}
while the expression for the massless gauge bosons $\gamma$ and $g$ reads
\begin{equation}
\left\{
    \begin{aligned}
       & \Gamma_{G_n  \rightarrow  \gamma  \gamma}=\frac{m_n^3}{80 \pi  \Lambda ^2}\\
       &\left. \right. \\
           & \Gamma_{G_n  \rightarrow  g  g}=\frac{m_n^3}{10 \pi  \Lambda ^2}\\
    \end{aligned}
    \right. \ .
\end{equation}
If the masses of the SM particles can be neglected this leads to combined width of
\begin{equation}
    \Gamma_{G_n \rightarrow \text{SM}}\approx \frac{73}{240\pi}\frac{m_n^3}{\Lambda^2}\ .
\end{equation}

The general expression for the decay into two lighter gravitons $G_m$ and $G_k$ is
\begin{align}
         & \Gamma_{G_n\rightarrow G_m G_k}=\frac{\chi _{\text{nkm}}^2}{17280 \pi  \Lambda^2}  \left[\left(m_k-m_m-m_n\right) \left(m_k+m_m-m_n\right) \left(m_k-m_m+m_n\right) \left(m_k+m_m+m_n\right)\right]{}^{5/2} \nonumber \\
         & \times  \left[26 m_k^6 \left(m_m^2+m_n^2\right)+14 m_k^4 \left(26 m_m^2 m_n^2+9 m_m^4+9 m_n^4\right) + 26 m_k^2 \left(14 m_m^4 m_n^2+14 m_m^2 m_n^4+m_m^6+m_n^6\right)\right. \nonumber\\
         & \left. +m_k^8+26 m_m^2 m_n^6+126 m_m^4 m_n^4 +26 m_m^6 m_n^2+m_m^8+m_n^8\right]/(m_k^4 m_m^4 m_n^7) \ ,
\end{align}
which simplifies if we consider two identical gravitons in the final state 
\begin{align}
          & \Gamma_{G_n  \rightarrow  G_m G_m}=\frac{\chi _{\text{nmm}}^2 \left(m_n^2-4 m_m^2\right){}^{5/2} \left(780 m_m^6 m_n^2+616 m_m^4 m_n^4+52 m_m^2 m_n^6+180 m_m^8+m_n^8\right)}{34560 \pi  \Lambda ^2 m_m^8 m_n^2} \ .
\end{align}
For a mixed graviton-radion final state we find
\begin{align}
           & \Gamma_{G_n  \rightarrow  r G_m}=\frac{ (k^2 e^{-2 \mu \pi }) \tilde{\chi} _{\text{nmr}}^2 \sqrt{\frac{\left(m_m^2+m_n^2-m_r^2\right){}^2}{4 m_n^2}-m_m^2} \left(\frac{\left(m_m^2+m_n^2-m_r^2\right){}^4}{m_m^4 m_n^4}+\frac{22 \left(m_m^2+m_n^2-m_r^2\right){}^2}{m_m^2 m_n^2}+76\right)}{240 \pi  \Lambda ^2 m_n^2} \ ,
           \end{align}
           while the width into a radion pair reads
\begin{align}
            & \Gamma_{G_n \rightarrow rr}=\frac{ \chi _{\text{nrr}}^2 \left(m_n^2-4 m_r^2\right){}^{5/2}}{960 \pi  \Lambda ^2 m_n^2}= \frac{ \left(m_n^2-4 m_r^2\right){}^{5/2}}{15 \pi  \Lambda ^2 m_n^2 \gamma_n^2}\ .
\end{align}
As long as $n$ is not very large these contributions are small compare to the width into SM particle and can be neglected.

\subsection{Radion Decay Widths}
The partial width into Higgs boson is given by
\begin{equation}
    \begin{aligned}
         &\Gamma_{r  \rightarrow  H H }=\frac{\sqrt{1-\frac{4 M_H^2}{m_r^2}} \left(2 M_H^2+m_r^2\right){}^2}{192 \pi  \Lambda ^2 m_r}\ .
    \end{aligned}
\end{equation}
For the decay into a Dirac fermion pair it reads
\begin{equation}
    \begin{aligned}
        &\Gamma_{r  \rightarrow \ \Bar{\psi}\psi}=N_c \frac{m_r m_{\psi }^2 \left(1-\frac{4 m_{\psi }^2}{m_r^2}\right)^{\frac{3}{2}} }{48 \pi  \Lambda ^2}\ .  
    \end{aligned}
\end{equation}
In the case of the massive gauge bosons $Z$ and $W^\pm$ the width is 
\begin{equation}
    \begin{aligned}
         &\Gamma_{r \ \rightarrow \ ZZ }=\frac{\sqrt{1-\frac{4 M_Z^2}{m_r^2}} \left(12 M_Z^4-4 m_r^2 M_Z^2+m_r^4\right)}{192 \pi  \Lambda ^2 m_r}\\
         &\left.\right. \\
        &\Gamma_{r  \rightarrow  W^+ \ W^- }=\frac{\sqrt{1-\frac{4 M_W^2}{m_r^2}} \left(12 M_W^4-4 m_r^2 M_W^2+m_r^4\right)}{96 \pi  \Lambda ^2 m_r}\\
    \end{aligned}
\end{equation}
while the decay into massless gauge bosons $\gamma$ and $g$ is given by
\begin{equation}
    \begin{aligned}
       &\Gamma_{r  \rightarrow  \gamma\gamma}=\frac{C_{\text{EM}}^2 \alpha _{\text{EM}}^2 m_r^3}{1536 \pi ^3 \Lambda ^2}\\
       &\left.\right. \\
        &\Gamma_{r  \rightarrow  g g}=\frac{C_S^2 m_r^3 \alpha _S^2}{192 \pi ^3 \Lambda ^2}\\
    \end{aligned}
\end{equation}
The radion does not couple to massless particles at tree-level. The loop contribution is encoded in the  
coefficients $C_{\text{EM}}$ and  $C_S$. These channels are not relevant to our analysis. A discussion of the coefficients can be found in \cite{Blum:2014jca}.
In principle, the radion can also decay to gravitons if it is heavy enough. In this work, the radion is assumed to be lighter than the first graviton and we do not include these decay channels.

\end{section}


\begin{section}{Coupling relations}
\label{app:sum}
For the simplification of the  high energy limit of amplitudes involving KK-gravitons the dimensionless coefficients 
\begin{equation}
    c_{n,\alpha} \equiv \sum\limits_{k}\chi_{nnk}\left(\frac{\gamma_k}{\gamma_n}\right)^\alpha \quad \text{,} \quad   d_{n,\alpha} \equiv \sum\limits_{k}\frac{\tilde{\chi}_{nkr}}{\gamma_n^2}\left(\frac{\gamma_k}{\gamma_n}\right)^\alpha \quad \text{and} \quad e_0=\sum\limits_{k}\chi_{krr}\ 
\end{equation}
play a central role.
The sum rules derived in \cite{deGiorgi:2020qlg} imply that
\begin{equation}
    c_{n,0}=1 \quad , \quad c_{n,2}=2\quad , \quad d_{n,-2}=1-\chi_{nrr} \ .
\end{equation}
In the main text we used in addition that
\begin{equation}
   \sum\limits_k \chi_{nmk}\gamma_k^4=\gamma_n^4+6\gamma_n^2\gamma_m^2+\gamma_m^4 \quad , \quad \sum\limits_{k}\tilde{\chi}_{nkr} = \gamma_n^2 \quad , \quad \sum\limits_{k}\chi_{krr}=1 \ ,
\end{equation}
and hence
\begin{equation}
    c_{n,4}=8 \quad , \quad  d_{n,0}=1 \quad , \quad e_0=1\ .
\end{equation}
to simplify the expressions of the cross section. 
In the following we provide an explicit proof for the interested reader.

As detailed in \cite{deGiorgi:2020qlg}, $c_{n,\alpha}$ and $d_{n,\alpha}$ can be evaluated with the help of the Fourier-Bessel expansion, see for example \cite{sneddon_1960}. If a function $f(x)$ is continuous on $[0, 1]$ such that $f(1)=0$, the integral
\begin{equation}
    \int\limits_0^1 \diff{x} \ x^{1/2} f(x)
\end{equation}
exists and is absolutely convergent, and $f(x)$ has limited total fluctuation, it can be expanded in series in terms of any Bessel-function $J_\nu$:
\begin{equation}
   f(x)= \sum\limits_{k=1}^\infty a_{\nu,k} J_\nu(\gamma_{\nu,k}x) \ ,
\end{equation}
where $\gamma_{\nu,k}$ is the $k-$th root of $J_\nu$ and
\begin{equation}
    a_{\nu,k}= \frac{2}{J_{\nu+1}(\gamma_{\nu,k})^2} \int\limits_0^1 \diff{u} \  u f(u) J_\nu(\gamma_{\nu,k} u) \ .
\end{equation}
Since we will work only with the roots of $J_1$, we define $\gamma_k$ as the $k-$th root of $J_1$ and we will set also $\nu=1$ such that
\begin{equation}
\label{fc-style}
    f(x)=2\sum\limits_{k=1}^\infty \frac{J_1(\gamma_k x)}{J_2(\gamma_k)^2}\int\limits_0^1 \diff{u} \  u f(u) J_1(\gamma_{k} u) \ .
    \end{equation}
    Our strategy consists of picking an appropriate function $f(x)$, such that we can obtain the relations of interest. We will also use some relations derived in \cite{deGiorgi:2020qlg} and the definitions of the couplings given in eq. \ref{def:couplings}.
\paragraph{Relation 1: $c_{n,4}$}
Let us recall the definition of
\begin{equation}
 \tilde{\chi}_{knm} \equiv -2\frac{\gamma_k \gamma_n}{J_0(\gamma_k)J_0(\gamma_n)J_0(\gamma_m)}\int\limits_0^1 \diff{u} \ u^3 J_1(\gamma_k u)J_1(\gamma_n u)J_2(\gamma_m u) \ ,
\end{equation}
which satisfies $\gamma_k^2\chi_{nmk}=\tilde{\chi}_{knm}+\tilde{\chi}_{kmn}$ \cite{deGiorgi:2020qlg}.
A convenient choice of $f(x)$ that satisfies $f(1)=0$ is 
\begin{equation}
\label{eq:rep1}
    f(x)=\frac{\gamma_n}{J_0(\gamma_n)J_0(\gamma_m)}J_1(\gamma_n x)J_2(\gamma_m x) x^2 \ .
\end{equation}
Using the fact that $J_0(\gamma_n)=-J_2(\gamma_n)$ and multiplying and dividing by $\gamma_k$, we have:
\begin{equation}
\label{eq:rep2}
    f(x)=\sum\limits_{k=1}^\infty \frac{J_1(\gamma_k x)}{\gamma_k J_2(\gamma_k)}\tilde{\chi}_{knm} \ .
\end{equation}
Differentiating both sides three times:
\begin{equation}
    f'''(x)=\sum\limits_k \tilde{\chi}_{knm}\frac{\left(\gamma_k ^2 x^2-3\right) J_2(x \gamma_k )}{x^2 J_2(\gamma_k )}\overset{x=1}{\rightarrow}\sum\limits_k \tilde{\chi}_{knm} (\gamma_k^2-3) \ .
\end{equation}
But on the other side:
\begin{equation}
    f'''(x)|_{x=1}=\gamma_n^2 \left(3 \gamma_m^2+\gamma_n^2-3\right) \ .
\end{equation}
We then have a very helpful relation that will be useful soon:
\begin{equation}
    \label{eq:mainsumrule}
    \sum\limits_k \tilde{\chi}_{knm}\gamma_k^2 =\gamma_n^4+3\gamma_n^2\gamma_m^2 \ .
\end{equation}
Combining this expression and $\gamma_k^2\chi_{nmk}=\tilde{\chi}_{knm}+\tilde{\chi}_{kmn}$  directly leads to
\begin{equation}
      \sum\limits_k \chi_{nmk} \gamma_k^4 =\left(\gamma_n^4+6\gamma_n^2\gamma_m^2+\gamma_m^4\right) \ .
\end{equation}
Using the case $m=n$, $c_{n,4}$ is thus
\begin{equation}
    c_{n,4}=\sum\limits_k \chi_{nnk} \left(\frac{\gamma_k}{\gamma_n}\right)^4=8 \ .
\end{equation}

\paragraph{Relation 2: $d_{n,0}$}
As before, the way to get the desired result is a good choice of $f(x)$; in this case we take
\begin{equation}
    f(x) = x^2 \frac{J_1(\gamma_n x)}{J_0(\gamma_n x)}\gamma_n \ ,
\end{equation}
from which follows
\begin{equation}
    f(x)=\sum\limits_k \frac{J_1(\gamma_k x)}{\gamma_k J_0(\gamma_k)}\tilde{\chi}_{knr} \ .
\end{equation}
Differentiating on both sides and evaluating the expressions at $x=1$
\begin{equation}
    \begin{split}
        f'(x)\left.\right|_{x=1}&=\sum\limits_k \tilde{\chi}_{knr}\\
        &=\gamma_n^2 \ .
    \end{split}
\end{equation}
Then the desired result follows
\begin{equation}
    \sum\limits_k \tilde{\chi}_{knr}=\gamma_n^2\ .
\end{equation}
Therefore we get
\begin{equation}
    d_{n,0} = 1 \ .
\end{equation}

\paragraph{Relation 3: $e_{0}$}
We focus now on the coefficient $\chi_{nrr}$, whose definition is given in eq. \ref{def:couplings}. It reads
\begin{equation}
    \chi_{nrr}=\frac{8}{\gamma_n^2} \ .
\end{equation}
In \cite{Rayleigh}, it is proved that  
\begin{equation}
    \sum\limits_{n=1}^\infty \frac{1}{\gamma_{\nu,n}^2} = \frac{1}{4(\nu+1)}\ ,
\end{equation}
where $\gamma_{\nu,n}$ is the $n$th-zero of the Bessel-$J_\nu$ function. In the case of interest,  $\nu=1$, this directly leads to $e_0=\sum\limits_{n=1}^\infty \chi_{nrr}=1$. 
\end{section}
\end{appendices}

\bibliographystyle{hieeetr}
\bibliography{extra_dim.bib}

\end{document}